\documentclass[prd,aps,twocolumn]{revtex4}
\usepackage[english]{babel}
\usepackage{amssymb,amsmath,amsfonts,amsbsy,epsfig,blindtext}
\usepackage{braket}
\usepackage{slashed}
\usepackage{bbm}

\usepackage{amsthm}
\usepackage{color}

\theoremstyle{definition}

\newcommand{\be}{\begin{equation}}
\newcommand{\ee}{\end{equation}}
\newcommand{\bea}{\begin{eqnarray}}
\newcommand{\eea}{\end{eqnarray}}

\def\L{{\mathcal L}}

\def\L{{\mathcal L}}
\def\H{{\mathcal H}}

\def\p{{\bf p}}
\def\q{{\bf q}}
\def\k{{\bf k}}
\def\r{{\bf r}}
\def\x{{\bf x}}
\def\y{{\bf y}}
\def\z{{\bf z}}

\begin{document}

\title{Probability Density Functionals in Perturbation Theory within the QFT in-in Formalism}
\date{27 March, 2019}

\author{
Bruno Scheihing Hitschfeld
}

\affiliation{
Grupo de Cosmolog\'ia y Astrof\'isica Te\'orica, Departamento de F\'{i}sica, FCFM, \mbox{Universidad de Chile}, Blanco Encalada 2008, Santiago, Chile.}

\begin{abstract} 
We delve into the \textit{in-in} formalism and derive general expressions for the computation of $n$-point functions for a self-interacting scalar field by means of a probability density functional (PDF). Even though in typical situations these formulae are not simpler than the standard approaches to Quantum Field Theory (QFT), we find that in specific situations, such as a first-order computation with a non-trivial potential, they can significantly reduce the effort needed to evaluate quantities of interest. The presented formulae may be readily generalized to other types of fields (e.g., fermionic Dirac fields), and to interacting theories that contain various types of fields. An analogous result for the \textit{in-out} formalism is presented using the previously employed strategies. Finally, we apply the obtained results to first-order computations and work out specific examples where we explore their limitations and possible prospects for future computations.

\end{abstract}

\maketitle

\section{Introduction}

Ostensibly, the hallmark computations of Quantum Field Theory (QFT) have been carried out using the \textit{in-out} formalism, that is, relying on the study of the $S$-matrix and its observable consequences. To wit, practically all physical predictions tested in particle accelerators, such as the Large Hadron Collider (LHC), are developed in this framework. However, this type of results assumes a certain degree of control over the experiment at hand, and to test them we typically have to repeat scattering processes a significant number of times. This is not feasible, for instance, when testing predictions for cosmological perturbations: we only have one universe to observe. Consequently, the computational tool that has proved most suitable in this type of context is the \textit{in-in} formalism~\cite{Schwinger:570,Schwinger:1960qe,Bakshi:1962dv, Bakshi:1963bn, Keldysh:1964ud, Jordan:1986ug, Weinberg:2005vy, Adshead:2009cb}, where only the initial condition of the system is specified, therefore making predictions without ever referring to the final state. The computable quantities within this formalism are correlations in a given time slice, evolving the system from the initial time up to the moment of interest.

On the other hand, computing observable quantities using the \textit{in-in} formalism has turned out to be lengthier than one might have hoped for. The existence of simple diagrammatic rules for \textit{in-out} computations, which dramatically reduce the time needed to carry out a computation, does not have an equally simple counterpart in this formalism. Of course, it must be noted that analog rules do exist (see for instance~\cite{Weinberg:2005vy}), but are somewhat non-trivial in comparison to those that describe scattering processes in a time-independent and homogeneous background. Having this in mind, we shall not focus much of our attention on diagrammatic rules, nor on how to derive them. Instead, we will explore how to deal with computations that require to consider a non-trivial structure in the interaction terms, beyond the usually studied cubic and quartic terms. This need arises, for example, in situations wherein there is a highly oscillatory potential in field space, which would be completely missed by the first three or four terms of its Taylor expansion. By neglecting higher-order terms, any features in the potential are effectively washed out, which actually may be of consequence to the probability distribution. 

At this point, we must note that when considering this kind of interactions, it could be argued that one would first need to establish the theory's renormalizability properties before proceeding to make sensible predictions by the means of perturbation theory. While this may be the usual, cautious way to go, most of the results we will find do not require to specify a renormalization scheme beforehand; thus, we do not explicitly address their implementation. On the same grounds, we neither attempt to solve this issue for arbitrary interactions. Instead, we expect that the apparent generality of our results lets us know whether, and how, renormalization becomes necessary. Similarly, should infinities emerge, the corresponding regularization procedure should also be guided by it. In any case, as it should become clear later on this paper, the usual approach using counterterms order by order in the perturbative parameters can be readily implemented.

In this work, we mainly concern ourselves with the derivation of probability density functionals (PDFs) in the presence of nontrivial self-interactions, which we treat perturbatively to all orders. PDFs are objects of interest within a vast range of areas within physics, going all the way from anharmonic crystals~\cite{Peierls:1,Brout:1,Zaslavskii:1} through Wave Turbulence~\cite{Nazarenko:B} and other topics in condensed matter physics to Cosmology, where, for instance, they can describe the matter density contrast in Large Scale Structure~\cite{Bernardeau:2001qr,Bernardeau:2013oda,Blas:2015qsi,Ivanov:2018lcg}, or the primordial distribution of curvature perturbations generated during Inflation~\cite{Starobinsky:1994bd,Chen:2018uul,Chen:2018brw, Chen:2017ryl}. Thus, the existence of general formulas for their derivation within a QFT framework can prove useful for present and future computations. Additionally, they may be useful outside the quantum arena, since the perturbative techniques commonly applied within QFT are also appropriate in weakly nonlinear classical field theories.

The present paper has been organized as follows: we begin in section~\ref{sec:setup} describing the \textit{in-in} formalism and introduce the main object of interest for the subsequent computations. In section~\ref{sec:pointPDF} we discuss the structure of the distribution function that allows to compute $n$-point functions at every order in perturbation theory, constituting the main result of this paper. We also briefly discuss the analogous result for the case of anti-commuting fields, written in terms of Grassmann variables. Section~\ref{sec:inout} presents a brief discussion about the standard approach to scattering experiments, the \textit{in-out} formalism, where we give the analogous expressions to what would be an $n$-point function and the distribution that generates them. In section~\ref{sec:disc-app} we delve into the specific case of first-order computations, writing down the explicit formulae for two PDFs that may be of interest. Finally, in section~\ref{sec:examples} we present a concrete example that illustrates how a nontrivial interaction expresses itself through a 1-point PDF to all orders, and also we discuss a field potential that seems to show promise of tractability with these tools. From there, we proceed to provide our concluding remarks in section~\ref{sec:conclusions}. We let the reader be aware that throughout this article we have chosen units such that $\hbar = 1$.

\section{Set up} \label{sec:setup}
In what follows we will study the dynamics of a self-interacting real scalar field \(\varphi\), which for concreteness we take to live inside a 4-dimensional spacetime (although this is not essential). The theory is described by a Lagrangian density 
\be \label{lagrangian}
\L= \L_{\text{free}} - V(\varphi, x).
\ee
Here \(x\) is the spacetime coordinate, and \(\L_{\text{free}}\) defines a free theory, of which we assume its solutions to be known in terms of mode functions \(\varphi_k(t)\):
\bea
\varphi(\x , t) &=& \int_k \, \hat \varphi (\k , t) \, e^{- i \k \cdot \x } \\  \hat \varphi (\k , t) &=& \varphi_k(t) a(\k) + \varphi_k^*(t) a^\dag(-\k) 
\eea
where we have introduced $\int_k \equiv (2\pi)^{-3} \int d^3 k$ as shorthand notation.
The operators $a(\k)$ and $a^\dag(\k)$ are creation and annihilation operators that satisfy the following commutation relations:
\be
\left[ a (\k) , a^{\dag} (\k') \right] = (2 \pi)^3 \delta^{(3)} (\k - \k').
\ee 
Equivalently, we may describe the complete theory through a hamiltonian density $\H= \H_{\text{free}} + V(\varphi, x)$. Either using this starting point or the lagrangian density~\eqref{lagrangian}, this setup is capable of describing a vast number of theories, including non-autonomous systems where there are explicit time dependences in the parameters of the theory. 
Having established the foundations, our attention will now be focused on describing the machinery we will use to solve the theory perturbatively.

We now proceed to quantize the system adopting the interaction picture framework. That is, the quantum field $\varphi$ is written as $\varphi (\x , t) = U^\dag (t)  \varphi_I (\x , t)  U (t)$,
where $\varphi_I (\x , t)$ is the interaction picture field and evolves as a field of the free theory. Explicitly, it is given by
\bea
\varphi_I(\x , t) &=& \int_k \, \hat \varphi_I (\k , t) \, e^{- i \k \cdot \x }  \\
\hat{\varphi}_I (\k , t) &=& \varphi_k(t) a(\k) + \varphi_k^*(t) a^{\dag}(-\k).
\eea

On the other hand, $U (t)$ is the time evolution operator in the interaction picture (sometimes dubbed as propagator), which is given by
\be \label{propagator}
U (t) = \text{T} \exp \left\{ - i \int^{t}_{t_0 + i\epsilon |t_0|} \!\!\!\!\!\!\!\!\! d t' H_I (t')  \right\} 
\ee
where $\text{T}$ is the time ordering symbol, instructing to place operators evaluated at later times at the left of the expression, and operators evaluated at earlier times at the right. Here we have incorporated a prescription to evaluate the integral, in the form of a positive infinitesimal quantity \(\epsilon\), which takes care of selecting the proper \textit{in} state when \(t_0 \to -\infty\)~\cite{Weinberg:1995mt}. This could also be implemented by adding an imaginary part to the argument of the interaction Hamiltonian $H_I$~\cite{Adshead:2009cb}. 

Clearly, the object that determines how temporal evolution takes place is $H_I$, which is given by the potential evaluated at the interaction picture fields,
\be \label{inter}
H_I(\tau) =  \int_{\x} V( \varphi_I(\x,t), \x, t),
\ee
where $\int_{\x} \equiv \int d^3 \x$. In order to deal with $H_I$, one of the methods we will consider is to make a Fourier expansion of the potential over field space and expand the exponential through its power series
\be \label{expansion}
\begin{split}
V(\varphi, \x, t) =& \, \frac{1}{2\pi} \int_{-\infty}^{\infty} d \gamma \tilde{V}(\gamma, \x, t) e^{-i \gamma \varphi} \\
 =& \, \frac{1}{2\pi} \int_{-\infty}^{\infty} d \gamma \tilde{V}(\gamma, \x, t) \sum_{m=0}^{\infty} \frac{(-i\gamma \varphi)^m}{m!},
\end{split}
\ee
which is formally possible since $\varphi$ is a hermitian field with real eigenvalues. This representation of the potential has also been utilized in other recent QFT results~\cite{Chebotarev:2018udi,Alexandrov:2017xmr} (referred therein as part of the construction of the $\mathcal{S}$-matrix in the Efimov representation, introduced earlier on in~\cite{Efimov:1977tw} and other previous works of the same author). In principle, either starting from here or from a Taylor expansion we will have to deal with an infinite number of vertices, with an arbitrary number of external legs. Although this expansion in Fourier modes is not essential to the final result, it turns out to be helpful in some derivations, being particularly convenient within a diagrammatic approach.  

Our first matter of interest will be to compute \(n\)-point functions for the \(\varphi\) field
\be
\langle \varphi(\x_1,t) ... \varphi(\x_n,t) \rangle = \langle U^\dag (t) \varphi_I(\x_1,t) ... \varphi_I(\x_n,t) U (t) \rangle.
\ee
Expanding the interaction picture propagator using the Dyson series one can readily write down
\begin{widetext}
\be \label{dyson}
\begin{split}
\langle \varphi(\x_1,t) ... \varphi(\x_n,t) \rangle =& \, \langle \sum_{N=0}^{\infty} (-i)^N \sum_{l=0}^{N} (-1)^l \int_{t_0 - i\epsilon |t_0|}^t \!\!\!\!\!\!\!\!\! dt_l ...
\int_{t_0 - i\epsilon |t_0|}^{t_2} \!\!\!\!\!\!\!\!\! dt_1 H_I(t_1) ... H_I(t_l) \\ & \times \varphi_I(\x_1,t) ... \varphi_I(\x_n,t) \int_{t_0 + i\epsilon |t_0|}^t \!\!\!\!\!\!\!\!\! dt_{l+1} ... \int_{t_0 + i\epsilon |t_0|}^{t_{N-1}} \!\!\!\!\!\!\!\!\! dt_N H_I(t_{l+1}) ... H_I(t_N) \rangle,
\end{split}
\ee
and thus we only need to evaluate expectation values of the form
\be
\langle H_I(t_1) ... H_I(t_l)   \varphi_I(\x_1,t) ... \varphi_I(\x_n,t) H_I(t_{l+1}) ... H_I(t_N) \rangle
\ee
\end{widetext}
where we have yet to specify the \textit{in} state. 

We construct these states using the creation and annihilation operators of the free theory, which we can write in terms of the interaction picture field operator and its conjugate canonical momentum field operator. This may be accomplished by inverting the relations that define the field observables in momentum space
\bea
 \hat \varphi_I (\k , t) &=& \varphi_k(t) a(\k) + \varphi_k^*(t) a^\dag(-\k) \\ 
 \Pi^\varphi_I (\k , t) &=& \dot{\varphi}_k(t) a(\k) + \dot{\varphi}_k^*(t) a^\dag(-\k), 
\eea
and then write down the operator of interest that specifies the \textit{in} state. For instance, one could write a superposition of one-particle states as
\be
\ket{\Phi} = \int_\x W(\x) \hat \varphi_I(\x) \! \ket{0},
\ee
or, more generally, for a superposition of multi-particle states,
\be
\ket{\Phi} = \sum_{n}  \left( \int_{\x_1} ... \int_{\x_n}  W_n(\x_1,...,\x_n) \hat \varphi_I(\x_1) ... \hat \varphi_I(\x_n) \right) \! \ket{0},
\ee
where the functions $\{W_n(\x_1,...,\x_n)\}_n$ characterize the state, with $n$ labelling the number of ``particles'' in each term of the sum. Here we have introduced the Fock vacuum $\ket{0}$, which is annihilated by the $a(\k)$ operators, i.e.,  $a(\k) \ket{0} = 0$, and supports a ladder of states upon acting on it with the $a^\dagger (\k)$ operators.

Finally, we note that because \(\Pi^{\varphi}_I(x) = d\varphi_I(x)/dt\), and the temporal derivative only affects the mode functions, we only need to compute
\begin{widetext}
\be \label{npoint}
\begin{split}
\bra{0} \! \varphi_I(\y_1, t_0) ... \varphi_I(\y_J, t_0) H_I(t_1) ... H_I(t_l)   \varphi_I(\x_1,t) ...
 \varphi_I(\x_n,t) H_I(t_{l+1}) ... H_I(t_N) \varphi_I(\y_1, t_0) ... \varphi_I(\y_J, t_0) \! \ket{0}
\end{split}
\ee
\end{widetext}
while keeping track of which field is to be differentiated with respect to time at \(t_0\), and finally summing the necessary terms to reconstruct the desired \textit{in} state, which will lie within the Fock space of the free theory vacuum. In this process, it is neater to let the positions $\y_i$ be different at each side of the inner product; thus we will first compute the correlation with no repeated positions in the fields, and take the corresponding limits at the end of the computation. Note that we may reconstruct the expectation value of other operators of interest by taking derivatives and linear combinations of the \(n\)-point correlation functions~\eqref{npoint} at time $t$. 
 
In the following subsections we proceed to outline the steps leading to the main result, while leaving most of the details to Appendices~\ref{sec:loops} and~\ref{sec:arbitrary-initial}. With the benefit of hindsight, we will appreciate that the results we will obtain are a direct consequence of Wick's theorem~\cite{Wick:1950ee}, and thus they do not rely on the particular representation of the expansion chosen to compute perturbations with the potential (Taylor series, Fourier series, etc.).

\section{The Probability Density Functional} \label{sec:pointPDF}

Since usually all of the relevant information to describe the theory can be stored within the path integral, it is natural to expect that we may write down an explicit functional that generates the $n$-point functions at any order in perturbation theory.

In this section we derive such a functional, which allows us to compute $n$-point functions at a given time slice $t$. We refer the interested reader to Appendix~\ref{sec:loops} for some of the technical details.

It will prove useful to write down the propagator of the free theory as a fundamental object, both in momentum and position space
\bea
\Delta(t,t',p) &\equiv& \varphi^{}_p(t)\varphi^*_p(t') \\  \sigma^2(t,t',r) &\equiv& \frac{1}{(2\pi)^3} \int d^3p \; \Delta(t,t',p) e^{i\p \cdot \r} \nonumber \\ &=& \braket{0|\varphi_I(t,\x) \varphi_I(t',\y) |0},
\eea
where $r = |\x - \y|$.

\subsection{The loop contributions}

Ultimately, it is the interacting term $V(\varphi, \r, t)$ that which will lead to non-trivial signatures, if any, in the spectrum of the field $\varphi$. Given that we are taking a perturbative approach, these signatures must be reflected solely through correlations, such as equation~\eqref{npoint}. If we expand the potential $V$ inside $H_I(\tau) =  \int_{\x} V( \varphi_I(\x,t), \x, t)$ as a power series on $\varphi_I$, each term in the expansion will ``interact'' with up to as many other spacetime positions as the power of the particular term. These interactions are usually represented with diagrams, joining ``outer legs'', which represent the fields that are used to construct the observed states (\textit{in} or \textit{out} states), and ``internal legs'' that arise from the interaction terms. Each of these connections gives a contribution of $\sigma^2(t,t',r)$ to the process under consideration, which we will call propagators or covariances depending on the context. Within all of these connections one can find contractions between two vertices (in this context, vertex is short for a spacetime position where an interaction potential is evaluated and the number of fields associated to it). As a result, it is possible to encounter ``closed circuits'', such as $\sigma^2(t_1,t_2,r_{12}) \sigma^2(t_2,t_3,r_{23}) \sigma^2(t_3,t_1,r_{31}) $. Diagrammatically, each propagator is represented with a line; hence these kinds of contributions are represented by a line segment that closes in itself. Therefore one calls them ``loops''.

To evaluate the correlation in equation~\eqref{dyson}, we will first address the fully interacting contribution to equation~\eqref{npoint}, i.e. that which connects all coordinates of the fields that define the \textit{in} state with interaction vertices. Put simply, for the moment we will not be interested in the contributions that arise directly from the free theory and can be factored out. Note that this is not equivalent to a fully connected contribution because this would require all external legs to interact with each other through the vertices, which would be reflected through an overall Dirac delta in momentum space (provided that the system is translationally invariant).

As is shown in Appendix~\ref{sec:loops}, the fully interacting $n$-point correlator contains the following loop structure as a factor:
\begin{widetext}
\be \label{npoint6}
\begin{split}
\bra{0} \! \varphi_I(\z_1, t_0) ... \varphi_I(\z_J, t_0) H_I(t_1) ... H_I(t_l)   \varphi_I(\x_1,t) ... \varphi_I(\x_n,t) H_I(t_{l+1}) ... H_I(t_N) \varphi_I(\y_1, t_0) ... \varphi_I(\y_J, t_0) \! \ket{0}_{FI} \\
\supset 
\int_{\r_1} \int_{\varphi_1} ... \int_{\r_N} \int_{\varphi_N} \dfrac{\partial^{n_1} V}{\partial \varphi_1^{n_1}} ... \dfrac{\partial^{n_N} V}{\partial \varphi_N^{n_N}}  \frac{\exp \left( -\frac{1}{2} \varphi_i \left(\Sigma_I^{-1}\right)_{ij} \varphi_j \right)}{\sqrt{(2\pi)^N |\text{det} \Sigma_I|}},
\end{split}
\ee
\end{widetext}
where we have omitted the propagators that connect the vertices with the outer legs. Here $n_l$ is defined as the number of ``legs'' at vertex $l$ of the perturbative expansion that are connected to the fields defining the \textit{in} state, and $\Sigma_I$ is a (complex) symmetric matrix that has the position space propagators connecting the vertices $\r_l$ as entries. This matrix plays the role of a covariance matrix, and so we will treat as such.

Let us appreciate an important aspect of this last result: it is an expectation value over a (multivariate) gaussian probability density function. With this in mind (and Wick's thoerem), we claim that
\begin{widetext}
\be \label{npointn}
\begin{split}
\bra{0} \! \varphi_I(\z_1, t_0) ... \varphi_I(\z_J, t_0) H_I(t_1) ... H_I(t_l)   \varphi_I(\x_1,t) ... \varphi_I(\x_n,t) H_I(t_{l+1}) ... H_I(t_N) \varphi_I(\y_1, t_0) ... \varphi_I(\y_J, t_0) \! \ket{0} = \int_{\varphi_{\z_1}} \!\!\!\!\!\! ... \! \int_{\varphi_{\z_J}}\!\!\!\!\!\!  ...  \\
\int_{\varphi_{\x_1}} \!\!\!\!\!\!  ... \! \int_{\varphi_{\x_N}} \!\int_{\varphi_{\y_1}}  \!\!\!\!\!\! ... \! \int_{\varphi_{\y_J}} \! \int_{\r_1} \int_{\varphi_1} ... \int_{\r_N} \int_{\varphi_N} V(\varphi_1, \r_1, t_1) ...  V(\varphi_N, \r_N, t_N)  \frac{\exp \left( -\frac{1}{2} \pmb{\varphi}^T \cdot \left(\pmb{\Sigma}^{-1}\right) \cdot \pmb{\varphi} \right)}{\sqrt{(2\pi)^{N+n+2J} |\text{det} \pmb{\Sigma} |}} \varphi_{\z_1} ... \varphi_{\z_J} \varphi_{\x_1} ... \varphi_{\x_n} \varphi_{\y_1} ... \varphi_{\y_J} 
\end{split}
\ee
where \(\pmb{\varphi}^T \equiv (\varphi_{\z_1} \,\,\, ... \,\,\, \varphi_{\z_J}  \,\,\, \varphi_{1} \,\,\, ... \,\,\, \varphi_l \,\,\, \varphi_{\x_1} \,\,\, ... \,\,\, \varphi_{\x_n} \,\,\, \varphi_{l+1} \,\,\, ... \,\,\, \varphi_N \,\,\, \varphi_{\y_1} \,\,\, ... \,\,\, \varphi_{\y_J}  )\), and \(\pmb{\Sigma}\) is the corresponding $(N+n+2J) \times (N+n+2J)$ covariance matrix.  
\end{widetext}  
The covariances in this matrix are the propagators between the fields' corresponding spacetime positions, and they have their respective temporal arguments ordered within the propagators' arguments as the fields are in the definition of $\pmb{\varphi}^T$. For example, the covariance relating $\varphi_{\z_a}$ and $\varphi_{b}$ is $\sigma^2(t_0,t_b,|\z_a-\r_b|)$, and the one relating $\varphi_{\x_i}$ with $\varphi_{b}$ would be $\sigma^2(t,t_b,|\x_i-\r_b|)$ if $b\geq l+1$, while it would be $\sigma^2(t_b,t,|\x_i-\r_b|)$ if $b \leq l$. We omit the dependence of \(\pmb{\Sigma}\) on $(N,l)$ to ease the notation. Also, the integrals $\int_{\varphi}$ are shorthand for $\int_{-\infty}^{\infty} d\varphi$.

Equation~\eqref{npointn} describes the full $n$-point function, including both the free theory contributions and the interacting ones. 
Note that the free theory pairings are given precisely by a Gaussian distribution as in~\eqref{npointn}, only without the $\varphi_i$ terms. Instances of these would be $\sigma^2(t_0,t_0,|\z_a-\y_b|)$ or $\sigma^2(t,t_0,|\x_i-\y_b|)$. Since it is fairly easy to check that these contractions also arise from this expression, we have obtained an indication that we are on the right track.

\subsection{A corollary of Wick's theorem}

We now proceed to prove the claim introduced in the previous section: let the potential $V$ be given by its Taylor expansion about $\varphi = 0$, with spacetime-dependent coefficients $c_m(\r,t)$
\be \label{taylorV}
V(\varphi,\r,t) = \sum_{m=0}^{\infty} \frac{c_m(\r,t)}{m!} \varphi^m,
\ee
and let us use this in~\eqref{npointn}. The right-hand side of the equation now reads
\begin{widetext}
\be
\begin{split} \label{proof1}
\int_{\r_1} ... \int_{\r_N} \sum_{m_1=0}^{\infty} ... \sum_{m_N=0}^{\infty} \frac{c_{m_1}(\r_1,t_1) ... c_{m_N}(\r_N,t_N)}{m_1! ... m_N!} \\ \times
 \int_{\varphi_{\z_1}} \!\!\!\!\!\! ... \! \int_{\varphi_{\z_J}} \int_{\varphi_{\x_1}} \!\!\!\!\!\!  ... \! \int_{\varphi_{\x_N}} \!\int_{\varphi_{\y_1}}  \!\!\!\!\!\! ... \! \int_{\varphi_{\y_J}} \!  \int_{\varphi_1} \int_{\varphi_N} \frac{\exp \left( -\frac{1}{2} \pmb{\varphi}^T \cdot \left(\pmb{\Sigma}^{-1}\right) \cdot \pmb{\varphi} \right)}{\sqrt{(2\pi)^{N+n+2J} |\text{det} \pmb{\Sigma} |}} \varphi_{\z_1} ... \varphi_{\z_J} \varphi_{\x_1} ... \varphi_{\x_n} \varphi_{\y_1} ... \varphi_{\y_J} \varphi_1^{m_1} ... \varphi_N^{m_N}.
\end{split}
\ee
 \end{widetext}
The second line in this last expression is nothing more than a moment of a multivariate gaussian distribution. Therefore, the result is the sum over all pairings of fields of the product of the corresponding covariances. On the other hand, if we go back to the starting point~\eqref{npoint}, we have
\begin{widetext}
\be \label{proof2}
\begin{split}
 \int_{\r_1} ... \int_{\r_N} \sum_{m_1=0}^{\infty} ... \sum_{m_N=0}^{\infty} \frac{c_{m_1}(\r_1,t_1) ... c_{m_N}(\r_N,t_N)}{m_1! ... m_N!}
 \bra{0} \! \varphi_I(\z_1, t_0) ... \varphi_I(\z_J, t_0) \varphi_I(\r_1, t_1)^{m_1} \\ ... \varphi_I(\r_l, t_l)^{m_l} \varphi_I(\x_1,t) ... \varphi_I(\x_n,t) \varphi_I(\r_{l+1}, t_{l+1})^{m_{l+1}} ... \varphi_I(\r_{N}, t_N)^{m_N} \varphi_I(\y_1, t_0) ... \varphi_I(\y_J, t_0) \! \ket{0}, 
\end{split}
\ee 
\end{widetext}
where the vacuum expectation value, per Wick's theorem, is exactly the sum, over all the possible pairings of fields, of the product of the free-theory two-point functions associated to the pairings, which in turn are exactly the covariances we have defined earlier. Hence the last two expressions are equal and therefore~\eqref{npointn} holds as written.

In order to connect this with the usual diagrammatic approach, note that in this last step the sum over all possible pairings is exactly what gives rise to propagators connecting vertices, and as may be seen from Appendix~\ref{sec:loops}, the flow of momenta through the diagrams appears by taking the Fourier transform to momentum space of each propagator. In this sense, we have only rewritten a known statement in an apparently more complicated manner. However, in this way it is possible to appreciate some aspects of perturbation theory that usually remain obscure in a diagrammatic approach. 

\subsection{An explicit result for the PDF of a self-interacting scalar field at every order} \label{subsec:main-PDF}

Now we return to equation~\eqref{dyson}, which is the object we want to characterize through a probability distribution. This PDF must be able to generate $n$-point functions to any order in perturbation theory, and should deliver a path integral in the formal limit $N \to \infty$. Furthermore, it should always be positive when reduced to a finite number of points at which to evaluate the field. The breakdown of this property would be a clear indicator that higher-order terms are required to give a meaningful result.

In what follows, we will write down explicit results taking the free theory vacuum $\ket{0}$ as the \textit{in} state (thus omitting $\varphi_\z$ and $\varphi_\y$ in $\pmb \varphi$), but it is straightforward to get a more general result, which we list in Appendix~\ref{sec:arbitrary-initial}. The reason for doing this is that the structure of the computation we wish to emphasize, namely the interacting terms, is already contained within the correlations that come out of this choice.

\begin{widetext}
Using equation~\eqref{npointn}, the $n$-point function for the $\varphi$ field is given by
\be
\begin{split}
\langle \varphi(\x_1,t) ... \varphi(\x_n,t) \rangle =&  \sum_{N=0}^{\infty} (-i)^N \sum_{l=0}^{N} (-1)^l \int_{t_0 - i\epsilon |t_0|}^t \!\!\!\!\!\!\!\!\! dt_l ...
\int_{t_0 - i\epsilon |t_0|}^{t_2} \!\!\!\!\!\!\!\!\! dt_1 \int_{t_0 + i\epsilon |t_0|}^t \!\!\!\!\!\!\!\!\! dt_{l+1} ... \int_{t_0 + i\epsilon |t_0|}^{t_{N-1}} \!\!\!\!\!\!\!\!\! dt_N \int_{\r_1} ... \int_{\r_N}  \\ & \times \int_{\varphi_{\x_1}} \!\!\!\!\!\!  ...  \int_{\varphi_{\x_N}} \!\int_{\varphi_1} \!\!\!  ... \int_{\varphi_N} V(\varphi_1, \r_1, t_1) ...  V(\varphi_N, \r_N, t_N)  \frac{\exp \left( -\frac{1}{2} \pmb{\varphi}^T \cdot \left(\pmb{\Sigma}^{-1}\right) \cdot \pmb{\varphi} \right)}{\sqrt{(2\pi)^{N+n} |\text{det} \pmb{\Sigma} |}} \varphi_{\x_1} ... \varphi_{\x_n}.
\end{split}
\ee
This correlation is a moment of the distribution
\be \label{PDF}
\begin{split}
\rho_{\varphi} = \sum_{N=0}^{\infty} (-i)^N \sum_{l=0}^{N} (-1)^l \int_{t_0 - i\epsilon |t_0|}^t \!\!\!\!\!\!\!\!\! dt_l ...
\int_{t_0 - i\epsilon |t_0|}^{t_2} \!\!\!\!\!\!\!\!\! dt_1 \int_{t_0 + i\epsilon |t_0|}^t \!\!\!\!\!\!\!\!\! dt_{l+1} ... \int_{t_0 + i\epsilon |t_0|}^{t_{N-1}} \!\!\!\!\!\!\!\!\! dt_N \int_{\r_1} ... \int_{\r_N}  \\ \times \!\int_{\varphi_1} \!\!\!  ... \int_{\varphi_N} V(\varphi_1, \r_1, t_1) ...  V(\varphi_N, \r_N, t_N)  \frac{\exp \left( -\frac{1}{2} \pmb{\varphi}^T \cdot \left(\pmb{\Sigma}^{-1}\right) \cdot \pmb{\varphi} \right)}{\sqrt{(2\pi)^{N+n} |\text{det} \pmb{\Sigma} |}},
\end{split}
\ee
\end{widetext}
which is, in its own right, a probability density function for the field $\varphi$. It is important to keep in mind that, even though it is not explicitly stated, the covariance matrix $\Sigma$ is different for each pair $(N,l)$, in the manner discussed between~\eqref{npointn} and~\eqref{taylorV}. We can summarize this by stating that the times that are integrated over a contour shifted by $-i\epsilon |t_0|$ always go to the left in the covariances, and that those with $+i\epsilon |t_0|$ always go to the right. When two times have the same imaginary component, the covariance has its arguments time-ordered if the integration is with $+i\epsilon |t_0|$, and anti-time-ordered if the integration goes with $-i\epsilon |t_0|$. Keeping this in mind, we may formally factor the distribution defined by the exponential out of the spacetime integrations, write it as a functional integral, and then further rearrange~\eqref{PDF} to get
\begin{widetext}
\be \label{Path-integral1}
\begin{split}
\rho_{\varphi} =& \int D\varphi \frac{\exp \left( - \frac{1}{2} \varphi \cdot \left( {\Sigma}^{-1} \right) \cdot \varphi \right)}{\sqrt{|\text{det} (2\pi \Sigma) |}} \sum_{l=0}^{\infty} (+i)^l \int_{t_0 - i\epsilon |t_0|}^t \!\!\!\!\!\!\!\!\! dt_l \int_{\r_l} V(\varphi(\r_l, t_l), \r_l, t_l) \,  ...
\int_{t_0 - i\epsilon |t_0|}^{t_2} \!\!\!\!\!\!\!\!\! dt_1 \int_{\r_1} V(\varphi(\r_1, t_1), \r_1, t_1) \\ & \times \sum_{N=0}^{\infty} (-i)^N   \int_{t_0 + i\epsilon |t_0|}^t \!\!\!\!\!\!\!\!\! dt_{l+1} \int_{\r_{l+1}} V(\varphi(\r_{l+1}, t_{l+1}), \r_{l+1}, t_{l+1}) \,  ... \int_{t_0 + i\epsilon |t_0|}^{t_{N+l-1}} \!\!\!\!\!\!\!\!\! dt_{N+l} \int_{\r_{N+l}}  V(\varphi(\r_{N+l}, t_{N+l}), \r_{N+l}, t_{N+l}),  
\end{split}
\ee
\end{widetext}
where we have to stress that the arguments of the fields inside the potential are there merely as a label for the functional integral to read; they are \textit{not} to be integrated over by the spacetime integrals right away.

In this last expression, $\varphi$ contains both ``internal'' (those in the arguments of the interaction $V$) and ``external'' fields (those that appear in the observable, characterized by the positions $\x_i$). 
One additional formal step gives the resummation of the Dyson series
\begin{widetext}
\be \label{Path-integral2}
\begin{split}
\rho_{\varphi} =& \int \! D\varphi_+ D\varphi_-  \exp \left\{ +i\int^{t}_{t_0 - i\epsilon |t_0|} \!\!\!\!\!\!\!\!\! d t' \int_{\r} V(\varphi_-(\r,t'),\r,t') \right\}  \! \\ & \times \frac{\exp \left( - \frac{1}{2} \varphi \cdot \left( {\Sigma}^{-1} \right) \cdot \varphi \right)}{\sqrt{|\text{det} (2\pi \Sigma) |}}   \exp \left\{ - i \int^{t}_{t_0 + i\epsilon |t_0|} \!\!\!\!\!\!\!\!\! d t' \int_{\r} V(\varphi_+(\r,t'),\r,t') \right\}.
\end{split}
\ee
\end{widetext}
Here we have made a distinction between the fields $\varphi_+$ and $\varphi_-$ (as is usually done in the Schwinger-Keldysh formalism) in order to be unambiguous regarding the order in which the covariances have their temporal arguments arranged: the $\varphi_+$ field always has its corresponding times to the right and time-ordered among themselves, while those of $\varphi_-$ always go to the left and anti-time-ordered among themselves. It is no longer necessary to write down the time ordering symbols, because the ordering prescription is already implemented through the definition of $\Sigma$. 
From this point one can derive an expression with more resemblance to the usual path integral formulation~\cite{Weinberg:2005vy}. It is also relevant to keep in mind that the matrix $\Sigma$ also has entries for external $\varphi$ fields, which are neither $\varphi_+$ nor $\varphi_-$: the Schwinger-Keldysh fields only account for the inner structure of the theory. To emphasize this, note that an expectation value for an observable quantity is computed by integrating over the corresponding external field variables 
\be
\langle f \rangle (\r_1, ..., \r_n ; t) = \int D\varphi \, \rho_{\varphi} \, f(\varphi(\r_1,t), ... ,\varphi(\r_n,t);t),
\ee
as one would expect.

A remark is in order here. Throughout this paper, when we write $\int D\varphi$ we mean to integrate over the range of eigenvalues of the field operator (heretofore the real line) for each spacetime position relevant in the integration, and reduce the corresponding (usually Gaussian) distribution to the relevant coordinates. For instance, in this last equation $\int D\varphi = \int_{\varphi(\r_1,t)} ... \int_{\varphi(\r_n,t)} $: the RHS is composed of a product of $n$ real integrals from $-\infty$ to $\infty$ for each field, thus sweeping over all of their possible eigenvalues.

The distribution $\rho_{\varphi}$ is normalized, in the sense that $\int D\varphi \rho_{\varphi} = 1$, as can be readily seen order by order from the perturbative expansion~\eqref{PDF}: if we disregard the $\epsilon$ prescription, then only $N=0$ gives a nonzero contribution, since by having integrated out the external fields, the argument of the time integrations is the same for every $l$, and what remains is a sum (with signs) over integration domains, which cancel out identically. Hence, for finite $t_0$ we have $\int D\varphi \rho_{\varphi} = 1$, and then $\lim_{t_0 \to -\infty} \int D\varphi \rho_{\varphi} = 1$, which is what is usually meant by $t_0 = -\infty$. Diagrammatically, this implies the cancellation of loop diagrams in the interacting theory that are disconnected from the external legs.

Me must note, though, that for computational purposes, equation~\eqref{Path-integral2} may be as useful as the starting point~\eqref{propagator} because ultimately both are formal expressions. However, it makes manifest one of the fundamental aspects of perturbation theory: the result is expressed only in terms of the quantities of the free theory, modulated by the perturbation $V$; no new propagators are introduced at a basic level.

As this section's final comment, we note that including self-interactions involving derivatives of the field is also feasible. However, in order to represent all the contractions through a Gaussian distribution in an unambiguous manner, it is desirable that the field and its conjugate momenta be ordered in a definite and uniform way within the Hamiltonian. If this is not the case, then one would probably be forced to add extra labels to the integrations so as to implement the different ordering prescriptions.

\subsection{The case of Dirac fields}

Analogously to what we have been doing so far, we may reconstruct a probability distribution functional for an interacting theory of anti-commuting fields. We consider a theory given by $\H= \H_{\text{free}} + V(\psi^{\alpha}, \bar{\psi}^{\beta}, x)$, where $\psi(x)$ is a Dirac spinor field. The relevant result for Gaussian expectation values when dealing with Grassmann variables is
\be
\langle A(\theta_a^{\alpha},\bar{\theta}_b^{\beta} ) \rangle = \int \left( \prod_{i,\iota} d \theta_i^{\iota} d\bar{\theta}_i^{\iota} \right) A(\theta_a^{\alpha}, \bar{\theta}_b^{\beta}) \frac{\exp\left( \bar{\theta}_i S^{-1}_{ij} \theta_j \right)}{\text{det} S^{-1}}.
\ee
Thus, it is possible to repeat the reasoning that led us to Eq.~\eqref{PDF} by expanding the ``potential'' $V$ as a polynomial of Grassmann variables (which actually is the very definition of $V$ in this context). In the last expression we introduced $S_{ij}$ as a matrix of complex-valued propagators between the corresponding Grassmann variables (we have omitted the spinor index in the sum $\bar{\theta}_i S^{-1}_{ij} \theta_j$). We use greek letters to denote spinor components and latin indices to number the interaction vertices. 

The only difference with our previous computations is that we are now faced with some numbers that do not commute. However, taking both the free and complete Hamiltonian (and therefore the interaction) to be commuting numbers, there is no problem in rearranging the positions of the potential $V$. Consequently, in an analogous manner we obtain
\begin{widetext}
\be \label{PDF-Grassmann}
\begin{split}
\rho_{\psi} = \sum_{N=0}^{\infty} (-i)^N \sum_{l=0}^{N} (-1)^l \int_{t_0 - i\epsilon |t_0|}^t \!\!\!\!\!\!\!\!\! dt_l ...
\int_{t_0 - i\epsilon |t_0|}^{t_2} \!\!\!\!\!\!\!\!\! dt_1 \int_{t_0 + i\epsilon |t_0|}^t \!\!\!\!\!\!\!\!\! dt_{l+1} ... \int_{t_0 + i\epsilon |t_0|}^{t_{N-1}} \!\!\!\!\!\!\!\!\! dt_N \int_{\r_1} ... \int_{\r_N}   \\  \times  \!\int  \left( \prod_{i,\iota} d \theta_i^{\iota} d\bar{\theta}_i^{\iota} \right) \frac{\exp\left( \bar{\theta}_i S^{-1}_{ij} \theta_j \right)}{\text{det} S^{-1}}  V(\theta_1^{\alpha}, \bar{\theta}_1^{\beta}, \r_1, t_1) ...  V(\theta_N^{\alpha}, \bar{\theta}_N^{\beta}, \r_N, t_N) .
\end{split}
\ee
\end{widetext}
Naturally, the integrals $\int d\theta$, $\int d\phi$ are no longer over the real line, but over complex Grassmann numbers. The spacetime-dependent covariance matrix $S$ follows the same conditions that $\Sigma$ does in the previous sections regarding the ordering of time coordinates within propagators, and is equally dependent on $(N,l)$. In very much the same way we did in the previous section, it proves possible to sum this back into a functional integral.

Generalizing these results to theories with both types of fields (bosonic and fermionic) is straightforward: write the (gaussian) free theory distribution as a product of those corresponding to each individual theory, and let the interaction terms $V(\varphi, \psi)$ connect them to form more complex processes.

\section{An analogous derivation for the in-out formalism} \label{sec:inout} 

We can also use the tools developed so far to write down expressions for the $S$ matrix. From~\cite{Weinberg:1995mt}, we have $S_{\beta \alpha} = \bra{\Phi_{\beta}} \! S \! \ket{\Phi_{ \alpha }}$, where $S = U(+\infty, -\infty)$, $U(t, t_0) \equiv e^{iH_0 t} e^{-i H(t - t_0)} e^{-i H_0 t_0} = \text{T} \exp \left( -i \int_{t_0}^{t} dt' \, H_I(t') \right)$, and $\ket{\Phi_{\alpha}}, \ket{\Phi_{\beta}}$ refer to the states of the free theory that are asymptotically equal to the relevant \textit{in} and \textit{out} states. In this context, we may write
\begin{equation}
\ket{\Phi_{\alpha}} = a^{\dagger}(\k_{\alpha_1}) ... a^{\dagger}(\k_{\alpha_n}) \! \ket{0},
\end{equation}
and therefore the object of interest is
\begin{equation} \label{in-outSa}
\begin{split}
\bra{0}\! a(\k_{\beta_1}) ... a(\k_{\beta_m}) \, & \text{T} \exp \left( -i \int_{-\infty}^{\infty} dt \, H_I(t) \right) \\ & \,\,\,\,\,\,\,\,\,\,\,\, \times a^{\dagger}(\k_{\alpha_1}) ... a^{\dagger}(\k_{\alpha_n}) \! \ket{0}.
\end{split}
\end{equation}
We omit the $\epsilon$ terms during this section, as they can be easily tracked and reinstated if necessary. In order to use our previous strategies, we may start by expressing the $a,a^{\dagger}$ operators in terms of the interaction picture fields at $t=0$. This choice is inconsequential to the construction of \textit{in}/\textit{out} states, as the evolution of the free theory will only modify the phases by which the $a,a^{\dagger}$ operators need to be multiplied in order to recover an amplitude of the form of~\eqref{in-outSa}. Thus, the object that we can readily compute and use to probe different \textit{in}/\textit{out} states is
\begin{equation} \label{in-outS}
\begin{split}
\braket{f | i} = \bra{0} & \varphi_I(\x_{\beta_1},0) ... \varphi_I(\x_{\beta_m},0)  \\ &   
\times \, \text{T} \exp \left( -i \int_{-\infty}^{\infty} dt \, H_I(t) \right) \\ & \times  \varphi_I(\x_{\alpha_1},0) ... \varphi_I(\x_{\alpha_n},0) \! \ket{0}.
\end{split}
\end{equation}

However, in contrast with the \textit{in-in} formalism, there is an issue regarding the loop diagrams that are disconnected from the external legs: in this situation there is no cancellation coming from other terms. Since these contributions typically give rise to infinities, the standard approach is to remove them by hand, because they appear as a normalization factor in the partition function. 
Equivalently, we can formally divide the amplitude $\braket{f | i}$ by $\braket{0| S |0}$, since it is this term that appears as an overall factor in the exact result for every computation, and accounts for all of the disconnected pieces that appear in a diagrammatic expansion.

If we now perform a perturbative expansion on $V$ (i.e. on its typical amplitude) in analogy with our previous computations, we obtain
\begin{widetext}
\begin{equation} \label{in-out-biPDF}
\begin{split}
\braket{f | i} =  \sum_{N=0}^{\infty} (-i)^N \!\!\!  \int_{\r_1} ... \int_{\r_N} \int_{\varphi_{\x_{\alpha_1}}} \!\!\! ... \int_{\varphi_{\x_{\alpha_1}}}  \!\int_{\varphi_{\x_{\beta_1}}} \!\!\!  ... \int_{\varphi_{\x_{\beta_m}}} \!\int_{\varphi_1} \!\!\!  ... \int_{\varphi_N} \varphi_{\x_{\alpha_1}} ... \varphi_{\x_{\alpha_n}} \varphi_{\x_{\beta_1}} ... \varphi_{\x_{\beta_m}}  \\ \times  \int_{-\infty}^{\infty} \!\!\!\!\! dt_{1} \, ... \int_{-\infty}^{t_{N-1}} \!\!\!\!\!\!\!\!\! dt_N   \frac{\exp \left( -\frac{1}{2} \pmb{\varphi}^T \cdot \left(\pmb{\Sigma}_N^{-1}\right) \cdot \pmb{\varphi} \right)}{\sqrt{(2\pi)^{N+n+m} |\text{det} \pmb{\Sigma}_N |}} V(\varphi_1, \r_1, t_1) ...  V(\varphi_N, \r_N, t_N), 
\end{split}
\end{equation}
\begin{equation}
\begin{split}
\braket{0| S |0} =  \sum_{N=0}^{\infty} (-i)^N  \!\!\! \int_{\r_1} ... \int_{\r_N} \int_{\varphi_1} \!\!\!  ... \int_{\varphi_N}  \int_{-\infty}^{\infty} \!\!\!\!\! dt_{1} \int_{-\infty}^{t_1} \!\!\!\!\! dt_{2} \, ... \int_{-\infty}^{t_{N-1}} \!\!\!\!\!\!\!\!\! dt_N \, \frac{\exp \left( -\frac{1}{2} \pmb{\varphi}_N^T \cdot \left(\pmb{\Sigma}_{I,N}^{-1}\right) \cdot \pmb{\varphi}_N \right)}{\sqrt{(2\pi)^{N} |\text{det} \pmb{\Sigma}_{I,N} |}} V(\varphi_1, \r_1, t_1) ...  V(\varphi_N, \r_l, t_N).
\end{split}
\end{equation}
\end{widetext}
In these expressions the covariance matrix $\pmb{ \Sigma}_N$ is a square matrix with side $N+n+m$, in which the field entries with a $\beta$ index always have their corresponding time coordinates in the first argument of $\sigma^2(t,t',r)$. Similarly, the field entries with an $\alpha$ index always have their respective time coordinates in the second argument. The field entries associated to the perturbative expansion have their corresponding time coordinates time-ordered with respect to each other, so that later times occupy the first entry of $\sigma^2(t,t',r)$. Analogously, $\pmb{ \Sigma}_{I,N}$ is an $N\times N$ square matrix, but only containing the covariance block associated to the $N$ vertices and no correlation to the initial or final state. Finally, $\pmb{\varphi}_N$ represents the field vector $\pmb{\varphi}$ with only the $\varphi_i$ entries, with $i=1,...,N$.

The previous expressions can be used at any order, expanding $\braket{0| S |0}^{-1}$ in a power series and keeping the terms of interest up to the desired order. As before, we may formally re-sum and write this result using functional integrals
\begin{widetext}
\be \label{Path-integralc3}
\begin{split}
\braket{f | i} = \int \! D\varphi  \frac{\exp \left( - \frac{1}{2} \varphi \cdot \left( {\Sigma}^{-1} \right) \cdot \varphi \right)}{\sqrt{|\text{det} (2\pi \Sigma) |}} \varphi_{\x_{\alpha_1}} ... \varphi_{\x_{\alpha_n}}  \exp \left\{ - i \int^{\infty}_{-\infty} \!\!\!\! d t \int_{\r} V(\varphi(\r,t),\r,t) \right\}   \varphi_{\x_{\beta_1}} ... \varphi_{\x_{\beta_m}} \\
\times \left[\int \! D\phi  \frac{\exp \left( - \frac{1}{2} \phi \cdot \left( {\Sigma_I}^{-1} \right) \cdot \phi \right)}{\sqrt{|\text{det} (2\pi \Sigma_I) |}} \!  \exp \left\{ - i \int^{\infty}_{-\infty} \!\!\!\! d t \int_{\r} V(\phi(\r,t),\r,t) \right\}  \right]^{-1},
\end{split}
\ee
or equivalently, one can also write equation~\eqref{Path-integralc3} as 
\be
\braket{f | i} = \int D\varphi_{\rm external} \, \varphi_{\x_{\alpha_1}} ... \varphi_{\x_{\alpha_n}} \varrho_{\varphi}  \,\varphi_{\x_{\beta_1}} ... \varphi_{\x_{\beta_m}},
\ee
where the distribution that generates the scattering amplitudes is
\be
\begin{split}
\varrho_{\varphi} =&  \int \! D\varphi_{\rm internal}  \frac{\exp \left( - \frac{1}{2} \varphi \cdot \left( {\Sigma}^{-1} \right) \cdot \varphi \right)}{\sqrt{|\text{det} (2\pi \Sigma) |}} \exp \left\{ - i \int^{\infty}_{-\infty} \!\!\!\! d t \int_{\r} V(\varphi(\r,t),\r,t) \right\}   \\
& \times \left[\int \! D\phi  \frac{\exp \left( - \frac{1}{2} \phi \cdot \left( {\Sigma_I}^{-1} \right) \cdot \phi \right)}{\sqrt{|\text{det} (2\pi \Sigma_I) |}}  \exp \left\{ - i \int^{\infty}_{-\infty} \!\!\!\! d t \int_{\r} V(\phi(\r,t),\r,t) \right\}  \right]^{-1}.
\end{split}
\ee
\end{widetext}
When reading this, one has to keep in mind that $\Sigma$ stores all correlations, including those between external and internal fields, which are here defined, respectively, as those used to construct the \textit{in} and \textit{out} states and those that arise from the interactions, while $\Sigma_I$ only contains the covariances of the internal fields. Additionally, $\varphi$ contains both $\varphi_{\rm external}$ and $\varphi_{\rm internal}$. This latter contribution can be thought as normalization to the scattering amplitude that removes the interactions that do not take part in the physical process (i.e. the disconnected loop contributions).

Let us stress that $\varrho_\varphi$ is not a probability distribution: it is merely a distribution which upon integration along a function of field variables yields scattering amplitudes. The normalization is conventional to set the corresponding partition function (the functional Fourier transform) to unity when the source currents are turned off. 

Finally, let us note that the time ordering symbols have been removed from the expression, as $\Sigma$ is defined so that it contains time-ordered (Feynman) propagators between the vertices, thus implementing the standard diagrammatic rules of QFT. As before, let us remark yet again that $\Sigma$ also contains propagators that connect the external fields with the interaction vertices, with their corresponding temporal orderings. Although we could have simplified the notation further, we keep $\sigma^2(t,t',r)$ as the fundamental object in order to facilitate comparison with the \textit{in-in} formalism. The familiar path integral approach to QFT can be readily compared with this result.

\section{Applications to first-order computations} \label{sec:disc-app}

In this section we list two generic quantities that can be derived from our previous results and may find a number of applications. The first considers the computation of a probability distribution function for a scalar field $\varphi$, starting from the free theory vacuum $\ket{0}$ as the \textit{in} state. The second provides a probability density for the mode amplitudes in momentum space, with the same initial conditions. They are complementary, as they quantify different aspects of the system's observable quantities, but ultimately encode the same information: the structure of the self-interaction.

Both results can be regarded as Born approximations, in the sense that they are valid provided the potential is of small amplitude and that their results are linear on the self-interaction. This fact may prove useful for future studies of systems that have statistics that are very close to a Gaussian distribution and probe small departures from Gaussianity, simply because linear operations are easier to handle and, potentially, to be inverted. This opens the door to obtaining information on the underlying potential directly, thus shedding light on the fundamental structure of the theory that describes the phenomenon at hand.

\subsection{A 1-point PDF}

A problem of interest is to describe the probability of a quantum field having a certain amplitude at a given spacetime position. Let us here take a situation in which the system is originally in the vacuum state of the free theory, and an interacting term is turned on from $t=t_0$ onwards. This allows us to omit the $\epsilon$ prescription used to select states in the asymptotic past.

If we take the term $N=1$ in equation~\eqref{PDF} and consider it as a deviation $\Delta \rho$ from a Gaussian distribution, we find
\be \label{rpsifull}
\begin{split}
& \Delta \rho(\varphi_1,...,\varphi_n;z) = \\ & \int_{\r} \int_{t_0}^t dt' \, 2\text{Im}  \left\{ \int d\phi  \frac{\exp\left(-\frac{1}{2} \pmb{\varphi}^T \cdot \pmb{\Sigma}^{-1} \cdot \pmb{\varphi} \right)}{\sqrt{(2\pi)^{n+1} |\text{det} \, {\pmb\Sigma|}}}  V\left(\varphi, \r, t' \right) \right\},
\end{split}
\ee
where \(\varphi_i\) represents the amplitude of the field at $(\r_i,t)$, which we do not write as \(\varphi(\r_i, t)\) in order to emphasize their being real variables. Furthermore, we have denoted \(\pmb{\varphi}^T =  (\phi \,\,\, \varphi_1 \,\,\, ... \,\,\, \varphi_n  )\), and \(\pmb{\Sigma}\) is the corresponding covariance matrix
\be \label{cov}
  \pmb{\Sigma}=
  \left[ {\begin{array}{cccc}
   \sigma^2(t',t',0) & \sigma^2(t,t',r_{01}) & \cdots & \sigma^2(t,t',r_{0n})\\
   \sigma^2(t,t',r_{01}) & \sigma^2(t,t,0) & \cdots & \sigma^2(t,t,r_{1n})\\
   \sigma^2(t,t',r_{02}) & \sigma^2(t,t,r_{12}) & \ddots & \sigma^2(t,t,r_{2n})\\
   \vdots & \vdots & \cdots & \vdots\\
   \sigma^2(t,t',r_{0n}) & \sigma^2(t,t,r_{1n})  & \cdots & \sigma^2(t,t,0)\\
  \end{array} } \right],
\ee
with $r_{ij} = |\r_i - \r_j|$ and $\r = \r_0$. Furthermore, if we take $\r_1 = ... = \r_n \equiv \x$, then the Gaussian measure effectively reduces itself to two field coordinates, one internal and another external. Consequently, the covariance matrix becomes a $2\times2$ matrix.

If we further let $r = |\r|$, we arrive at

\begin{widetext}
\be \label{rpsi2}
\begin{split}
\rho(\varphi,\x,t) = \frac{e^{-\frac{\varphi^2}{2\sigma^2(t,t,0)}}}{\sqrt{2\pi \sigma^2(t,t,0)}} \left[ 1 + \int_{\r} \int_{t_0}^t dt'  \int_{-\infty}^{\infty} \!\!\!\!\! d\phi \,\, 2 \, \text{Im}  \left\{  \frac{e^{-\frac{(\phi - R(t,t',r)\varphi)^2}{2\kappa^2(t,t',r)}}}{\sqrt{2\pi \kappa^2(t,t',r)}} \right\} V\left(\phi, \r + \x, t' \right) \right].
\end{split}
\ee
\end{widetext}
where we have defined
\bea
\kappa^2(t,t',r) &\equiv& \frac{\sigma^2(t,t,0) \sigma^2(t',t',0)-\sigma^4(t,t',r)}{\sigma^2(t,t,0)} \\ R(t,t',r) &\equiv& \frac{\sigma^2(t,t',r)}{\sigma^2(t,t,0)}.
\eea

Equation~\eqref{rpsi2} is the promised result: a 1-point PDF for the scalar field at $(\x,t)$. In general, this can be evaluated numerically in a straightforward manner, typically once the (co)variances have been already regularized. However, if we had access to solving the integral over $t'$ at the $n$-point function level, the reconstruction of the PDF should even be more revealing of the underlying physics (see~\cite{Chen:2018uul, Chen:2018brw} for a realization of this type of analysis). 

In fact, if both the potential and the single-point variance of the field $\sigma^2(t,t,0)$ are independent of the spacetime coordinates, it is possible to retrieve information of the self-interaction directly from the connected $n$-point functions in terms of a Hermite polynomial expansion, and ultimately, to reconstruct the potential. This is a consequence of equation~\eqref{npointk} in the following section.

\subsection{A $k$-space PDF for the amplitude of the fluctuations}

In order to get a PDF in momentum space, we first need to determine the structure of the $n$-point functions in this representation. To that end, we first write down the fully connected $n$-point function explicitly at first order in $V$ and time $t$, which, to this order in the perturbation, is equal to the fully interacting contribution:
\begin{widetext}
\be \label{ndpoint}
\langle \varphi_{\r_1,...,\r_n}^n \rangle_c = \int_{t_0}^t dt' \int_{-\infty}^{\infty} \!\!\!\! d\varphi \frac{e^{-\frac{\varphi^2}{2\sigma^2(t',t',0)}}}{\sqrt{2\pi \sigma^2(t',t',0)}}
 \frac{\partial^n V}{\partial \varphi^n}\left(\varphi,\r,t' \right) \int_{\r} 2 {\rm Im} \left\{ \sigma^2(t,t',|\r - \r_1|) \;...\; \sigma^2(t,t',|\r-\r_n|) \right\}.
\ee
Integrating by parts over $\varphi$ and taking a Fourier transform to momentum space, we get
\be \label{npointk}
\langle \varphi_{\k_1,...,\k_n}^n \rangle_c = \int_{t_0}^t dt' \int_{-\infty}^{\infty} \!\!\!\! d\varphi \frac{e^{-\frac{\varphi^2}{2\sigma^2(t',t',0)}}}{\sqrt{2\pi \sigma^2(t',t',0)}} {\rm He}_n\left(\frac{\psi}{\sigma(t',t',0)} \right) \! V \! \left(\varphi,\r,t' \right) \int_{\r} 2 {\rm Im} \left\{ \frac{\Delta(t,t',k_1) e^{i\k_1\cdot \r}}{\sigma(t',t',0)} \;...\;  \frac{\Delta(t,t',k_n) e^{i\k_n\cdot \r}}{\sigma(t',t',0)}  \right\}.
\ee
\end{widetext}
On the other hand, when one thinks about measuring different modes of a given field, one has to take into account the experimental limitations. To that end, we define
\be
\bar{ \varphi}_{\k} \equiv \frac{3}{4\pi k_{\rm IR}^3} \int_{|\q - \k| < k_{\rm IR}} \!\!\!\!\!\!\!\!\!\!\!\!\!\!\!\!\!\!\! d^3 q \, \varphi_{\q}
\ee
where $k_{\rm IR}$ is an infrared cutoff, or a coarse-graining, that accounts for our not being able to measure arbitrarily large length scales (note that there is no $(2\pi)^{-3}$ factor beside the integral).

In the case of a quadratic theory, all of the relevant information is contained within the two-point function, which is also called the \textit{Power Spectrum}. Presently, it is given by
\be
\langle \bar{\varphi}_\k \bar{\varphi}_{-\k} \rangle_{\rm free} =  \frac{(2\pi)^3}{\left(\frac{4\pi}{3} k_{\rm IR}^3 \right)^2} \int_{|\q - \k| < k_{\rm IR}} \!\!\!\!\!\!\!\!\!\!\!\!\!\!\!\!\!\!\! d^3 q \, \Delta(t,t,q).
\ee
which, depending on the background metric, may be time-dependent. However, in the presence of an interaction term, the theory no longer has Gaussian statistics, and consequently the amplitude of the modes is no longer determined only through the two-point function.

For our present purposes, since the amplitude of a mode can be characterized by $|\varphi_\k|^2 = \varphi_\k \varphi_\k^{\dagger} = \varphi_\k \varphi_{-\k}$, the natural quantity to try and compute is
\be \label{fullnpointk}
\langle (\bar{\varphi}_\k \bar{\varphi}_{-\k})^n \rangle = \sum_{m=0}^{n} \frac{n!^2 \langle \bar{\varphi}_\k \bar{\varphi}_{-\k} \rangle_{\rm free}^{n-m}}{m!^2 (n-m)!}  \langle (\bar{\varphi}_\k \bar{\varphi}_{-\k})^m \rangle_{c},
\ee
where the combinatorial factor arises from the counting of all possible contractions to form diagrams with $n$ external momenta evaluated at $\k$ and an additional set of $n$ momenta evaluated at $-\k$. Notice that~\eqref{fullnpointk} requires $|\k| > k_{\rm IR}/2$ for consistency, so that the free theory contractions can only join $\k$ with $-\k$.

\begin{widetext}
If we define
\be
\mathcal{F}_{n}[V;\r,t] \equiv \int_{-\infty}^{\infty} \!\!\!\! d\varphi \, \frac{e^{-\frac{\varphi^2}{2\sigma^2(t,t,0)}}}{\sqrt{2\pi \sigma^2(t,t,0)}}
{\rm He}_n\left(\frac{\varphi}{\sigma(t,t,0)} \right)  V\left(\varphi,\r,t \right),
\ee
we can readily write down the fully connected contributions
\be
\langle (\bar{\varphi}_\k \bar{\varphi}_{-\k})^n \rangle_{c} = \int_{t_0}^t dt' \int_\r \mathcal{F}_{2n}[V;\r,t'] \,  2 \text{Im} \left\{ \left( \frac{1}{(\frac{4\pi}{3} k_{\rm IR}^3)^2}  \int_{q,q' < k_{\rm IR}} \!\!\!\!\!\! \frac{\Delta(t,t',|\k-\q|) \Delta(t,t',|\k-\q'|)}{\sigma^2(t',t',0)} e^{i(\q - \q') \cdot \r}  \right)^n  \right\},
\ee
and thus what remains is a problem of finding the PDF that generates~\eqref{fullnpointk}. In essence, we want a distribution $\mathcal{K}$ such that $\int d(|\bar{\varphi}_\k|^2) \mathcal{K}_{\k}(|\bar{\varphi}_\k|^2) (|\bar{\varphi}_\k|^2)^n = \langle (\bar{\varphi}_\k \bar{\varphi}_{-\k})^n\rangle $. 

In the interests of notational simplicity, let us define
\be
y = y(t,t',\r;\k) \equiv \frac{1}{(\frac{4\pi}{3} k_{\rm IR}^3)^2 \langle \bar{\varphi}_\k \bar{\varphi}_{-\k} \rangle_{\rm free}} \int_{q,q' < k_{\rm IR}} \!\!\!\!\!\! \frac{\Delta(t,t',|\k-\q|) \Delta(t,t',|\k-\q'|)}{ \sigma^2(t',t',0)} e^{i(\q - \q') \cdot \r}, \,\,\,\, \& \,\,\,\, x \equiv \frac{|\bar{\varphi}_\k|^2}{\langle \bar{\varphi}_\k \bar{\varphi}_{-\k} \rangle_{\rm free}}
\ee
\end{widetext}
where $\langle \bar{\varphi}_\k \bar{\varphi}_{-\k} \rangle_{\rm free}$ is the two-point function of the free theory.
Furthermore, we define a functional $T[V;\r,t](x)$ by
\be
\begin{split}
\int_0^{\infty} dx \, e^{-x} L_n\left(x \right) T[V;\r,t](x) = \frac{(-1)^n}{n!} \mathcal{F}_{2n}[V;\r,t].
\end{split}
\ee
where $L_n$ is the $n$-th Laguerre polynomial. This definition is always possible provided that $V(\varphi,\cdot,\cdot)$ be square integrable with respect to the Gaussian measure with variance $\sigma^2(t,t,0)$ because both sides of the expansion define the coefficients of square integrable functions in their respective Hilbert spaces: $(\{{\rm He}_n(x)\}_n, \frac{e^{-x^2/2}}{2\pi})$ and $(\{L_n(x)\}_n, e^{-x})$.

Then, it follows from the preceding definitions and some functional-algebraic manipulations that
\begin{widetext}
\be
\mathcal{K}_{\k}(x) = e^{-x} \left[1 + \int^t_{t_0} dt' \int_\r \int_{0}^{\infty} \!\! dz \, \frac{e^{ -\frac{xy+z}{1-y} }}{1-y} I_0\left( \frac{2\sqrt{xyz}}{1-y} \right) T[V;\r,t'](z) \right]
\ee
where $I_0$ is a modified Bessel function of the first kind, and
\be
T[V;\r,t](z) = \int_{-\infty}^{\infty} \!\!\!\! d\varphi \, \frac{e^{-\frac{\varphi^2}{2\sigma^2(t,t,0)}}}{\sqrt{2\pi \sigma^2(t,t,0)}}
 V\left(\varphi,\r,t \right) \frac{1}{2\pi i} \int_{\mathcal{C}} \frac{dv}{v} \exp \left( \frac{v\varphi}{\sigma(t,t,0)} - \frac{v^2}{2} + \frac{z}{v^2+1} \right),
\ee
with $\mathcal{C}$ a counterclockwise integration contour encircling the three singularities of the integrand: $0,+i,-i$. Finally, we may write the PDF in terms of $| \bar{\varphi}_\k|$
\be \label{k-PDF}
\mathcal{K}_{\k}(| \bar{\varphi}_\k|) = \frac{ 2 | \bar{\varphi}_\k| e^{-\frac{| \bar{\varphi}_\k|^2}{\langle \bar{\varphi}_\k \bar{\varphi}_{-\k} \rangle_{\rm free}} }}{ \langle \bar{\varphi}_\k \bar{\varphi}_{-\k} \rangle_{\rm free} } \left[1 + \int^t_{t_0} dt' \int_\r \int_{0}^{\infty} \!\! dz \, \frac{e^{ -\frac{y| \bar{\varphi}_\k|^2 +z \langle \bar{\varphi}_\k \bar{\varphi}_{-\k} \rangle_{\rm free}}{(1-y) \langle \bar{\varphi}_\k \bar{\varphi}_{-\k} \rangle_{\rm free}} }}{1-y} I_0\left( \frac{2 |\varphi_{\k}|}{(1-y) } \sqrt{\frac{yz}{\langle \bar{\varphi}_\k \bar{\varphi}_{-\k} \rangle_{\rm free}}} \right) T[V;\r,t'](z) \right]
\ee
as a distribution over $|\varphi_{\k}| \in (0,\infty)$.
\end{widetext}

This result may find applications, for instance, when generating initial conditions for the evolution of the universe after inflation, or even probing the landscape potential that generated those initial conditions through CMB or LSS statistics, in an analogous manner to what was done in~\cite{Chen:2018uul, Chen:2018brw} for the CMB. This result has both advantages and disadvantages over the approach implemented in those works. On the one hand, because all the modes in $k$-space are statistically independent, the result will be subject to far less intrinsic noise. But on the other hand, its analytical expression is more cumbersome and it will presumably require more data from smaller scales on the sky, as it only would be able to reconstruct a PDF (assuming isotropy) by counting occurrences of the fluctuations' amplitude over a sphere at fixed $|\k|$. 

As the final note of this section, it is worth mentioning that given a reconstruction of the $k$-space PDF $\mathcal{K}(x)$ from actual data, one can recover information about the even Hermite moments of the self-interaction $\mathcal{F}_{2n}$ through
\be
\begin{split}
\int_0^\infty & \!\! dx L_n(x) \mathcal{K}_{\k}(x) \\
& = \int_{t_0}^t dt' \int_{\r}  \mathcal{F}_{2n}[V;\r,t'] \, 2 {\rm Im} \left\{ y^n(t,t',\r;\k) \right\},
\end{split}
\ee
which, in the same spirit of last section's conclusions, is readily useful if $\mathcal{F}_{2n}[V;\r,t']$ does not depend on $\r$ nor $t'$. It must be noted that the PDF~\eqref{k-PDF} only contains information on the even part of the potential $V$, and therefore another complementary observable should be used to obtain information on the odd part of the potential.

This concludes our discussion on first-order results.

\section{Exploring computational prospects} \label{sec:examples}

In this section we study two examples where it is possible to make some analytical progress starting from the general expressions~\eqref{PDF} and~\eqref{in-out-biPDF}. The first consists of a localized self-interaction in spacetime, where we proceed to examine the limitations of the perturbative approach by comparing the first-order result with a numerical computation following the exact result. The second considers a specific type of self-interaction present at all points in spacetime, where we are able to solve the field integrals exactly, as a means of showing where the typical ``perturbative'' parameter will appear and what structure emerges at each order in perturbation theory.

\subsection{A concrete example} \label{sec:example}

To substantiate the formulae pushed forward earlier, let us examine a specific example in which we are able to obtain (numerically) an exact probability density function for a quantum field in presence of a nontrivial interaction, while taking the initial state to be the free theory vacuum $\ket{0}$. 

To keep matters as simple as possible, we will choose a potential that is easy enough to handle upon insertion in equation~\eqref{PDF}, but still nontrivial in structure regarding its field coordinate. Our choice for a self-interaction will be
\be
V(\varphi,\r,t) = \delta(t) \delta^{(3)}(\r) \mathcal{V} (\varphi),
\ee
leaving $\mathcal{V}(\varphi)$ to be chosen later on. It is worth mentioning that $\mathcal{V}$ is a dimensionless function, as the Dirac deltas account for the action integral's dimensionality.

We can think of this potential $V$ as a localized event in spacetime, or a bounded region of spacetime whose characteristic size is much smaller than the typical scale of variation of the propagator $\sigma^2$, where the scalar field of interest was subject to nontrivial interactions, encoded in the function $\mathcal{V}$. The advantage of having made this choice is that the subsequent results can be obtained independently of the background metric, whose information will be encoded within the propagator/covariance of the free field.
Replacing this self-interaction into~\eqref{PDF} and performing the spacetime integrals gives
\be
\begin{split}
\rho_{\varphi} = \sum_{N=0}^{\infty} (-i)^N \sum_{l=0}^{N} \frac{(-1)^l}{l! (N-l)!} & \int_{\varphi_+} \!\! \int_{\varphi_-} \!\! \mathcal{V}(\varphi_-)^l  \mathcal{V}(\varphi_+)^{N-l} \\ \times  &  \frac{\exp \left( -\frac{1}{2} \pmb{\varphi}^T \cdot \left(\pmb{\Sigma}^{-1}\right) \cdot \pmb{\varphi} \right)}{\sqrt{(2\pi)^{3} |\text{det} \pmb{\Sigma} |}}
\end{split}
\ee
where the integrals over the field variables have been reduced to two as the position of the interaction vertices collapsed into a unique location $(\r,t) = 0$. This series may be summed back into
\be \label{0dim-PDF}
\rho_\x(\varphi) = \int_{-\infty}^{\infty} \!\!\!\! d\varphi_+ \int_{-\infty}^{\infty} \!\!\!\! d\varphi_- e^{i \left( \mathcal{V}(\varphi_-) - \mathcal{V}(\varphi_+) \right)  } \frac{e^{\left(-\frac{1}{2} \pmb{\varphi}^T \cdot \pmb{\Sigma}^{-1} \cdot \pmb{\varphi} \right)}}{\sqrt{(2\pi)^{3} |\text{det} \, {\pmb\Sigma|}}} 
\ee
where $\pmb{\varphi}^T = (\varphi, \varphi_+, \varphi_- )$ is the field vector (with real entries) analogous to the previous instances of $\pmb{\varphi}^T$ in this work, and $\Sigma$ is the corresponding covariance matrix:
\be 
  \pmb{\Sigma}=
  \left[ {\begin{array}{ccc}
   \sigma^2(x,x) & \sigma^2(x,0) & \sigma^2(0,x) \\
   \sigma^2(x,0) & \sigma^2(0,0) & \sigma^2(0,0) \\
   \sigma^2(0,x) & \sigma^2(0,0) & \sigma^2(0,0) \\
  \end{array} } \right]
\ee
where we have written $\sigma^2(0,0)$ and $\sigma^2(x,x)$ as the variances of the field $\varphi$ at the coordinate origin (the point where the potential is active) and at the point where the field PDF is evaluated $x = (\r,t)$, respectively. Similarly, $\sigma^2(0,x)$ and $\sigma^2(x,0)$ are the covariances relating the coordinates $x$ and $0$, with their temporal arguments ordered so as to account for the time ordering in the formalism. 

\begin{figure}[t!]
\includegraphics[width=8.6cm]{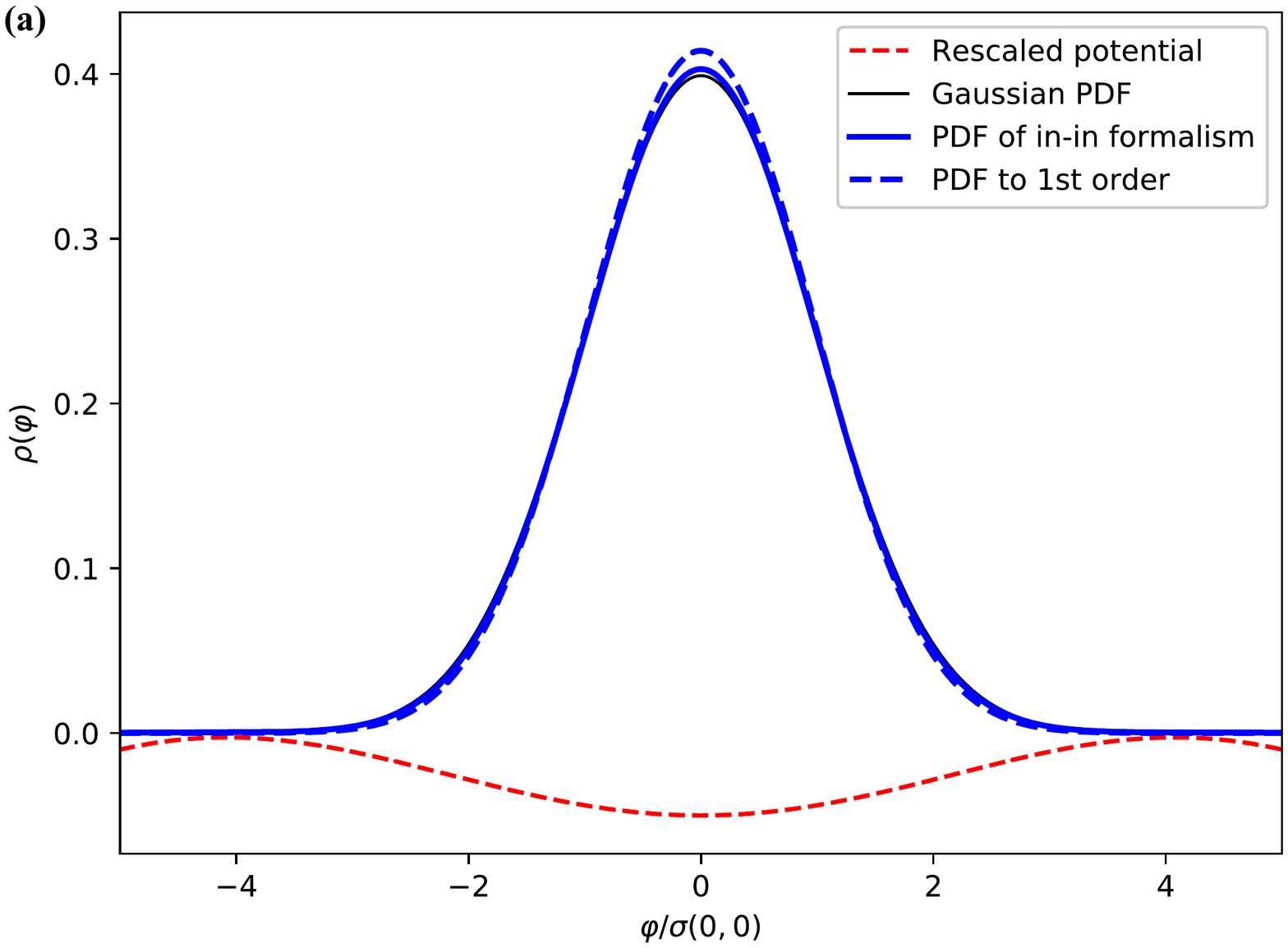}
\includegraphics[width=8.6cm]{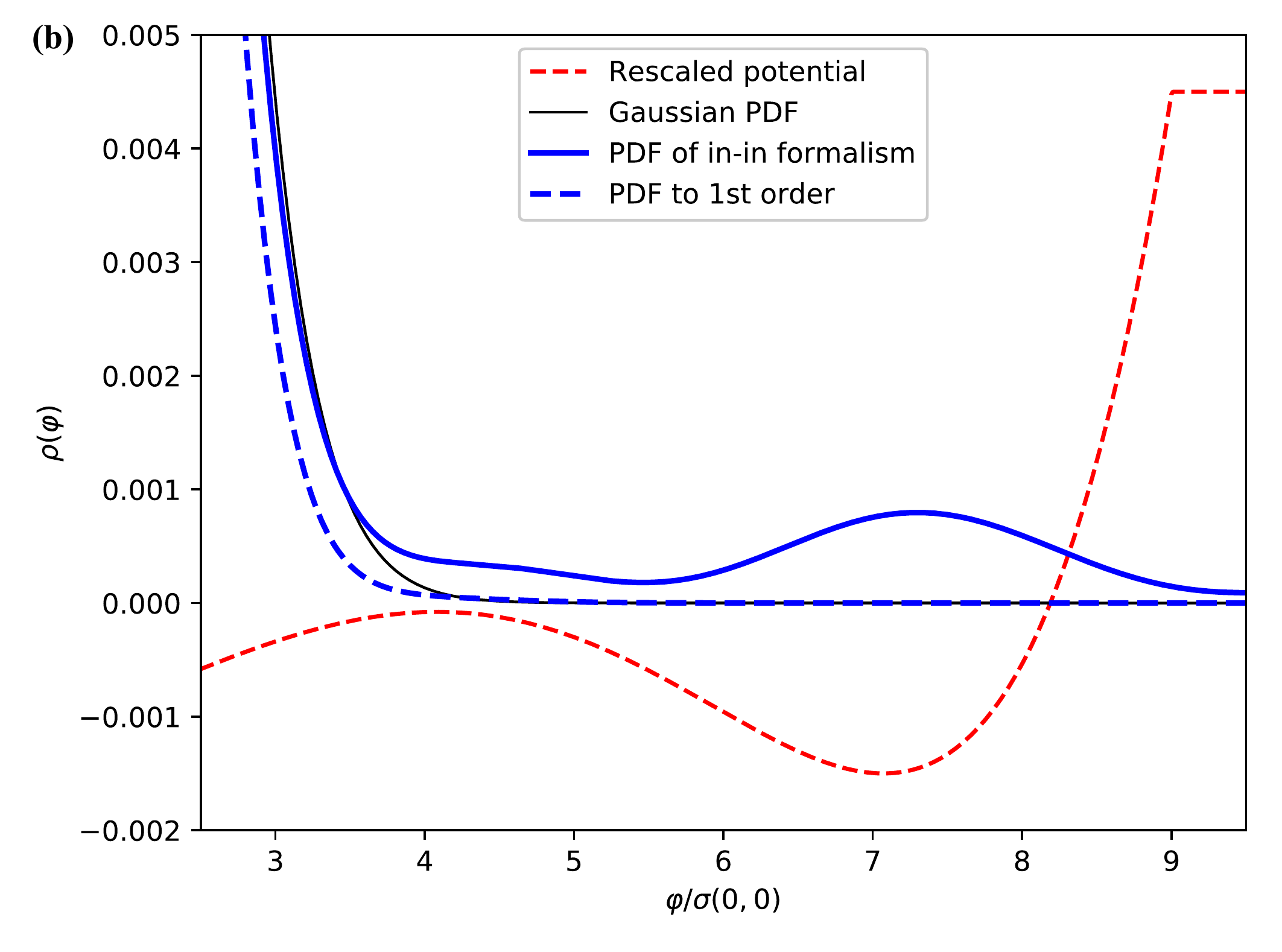}
\caption{The figure illustrates a typical PDF (solid blue) that may be obtained exactly using the \textit{in-in} formalism and the tools developed herein, in the presumably perturbative regime $A = 0.1$ of~\eqref{V-example}. A Gaussian PDF (solid black) and the first order result in perturbation theory (dashed blue) are shown for comparison. The PDF was generated with $\sigma^2(x,0) = (1/5 - 4i/5) \sigma^2(0,0) $. The potential that gives rise to the non-Gaussian deformations is also displayed for comparison (dashed red). Its amplitude is not to scale, but the field range has been rescaled by ${\rm Re}\{\sigma^2(x,0)\}/\sigma^2(0,0)$ in order to properly account for the transfer. (a) Shows the central region of the distribution $\varphi \in (\, -5\sigma(0,0) \, , \, 5\sigma(0,0) \, )$, and (b) displays the tail $\varphi \in (\, 2\sigma(0,0) \, , \, 10\sigma(0,0) \, )$.}
\label{fig:FIG_1}
\end{figure}

\begin{figure}[t!]
\includegraphics[width=8.6cm]{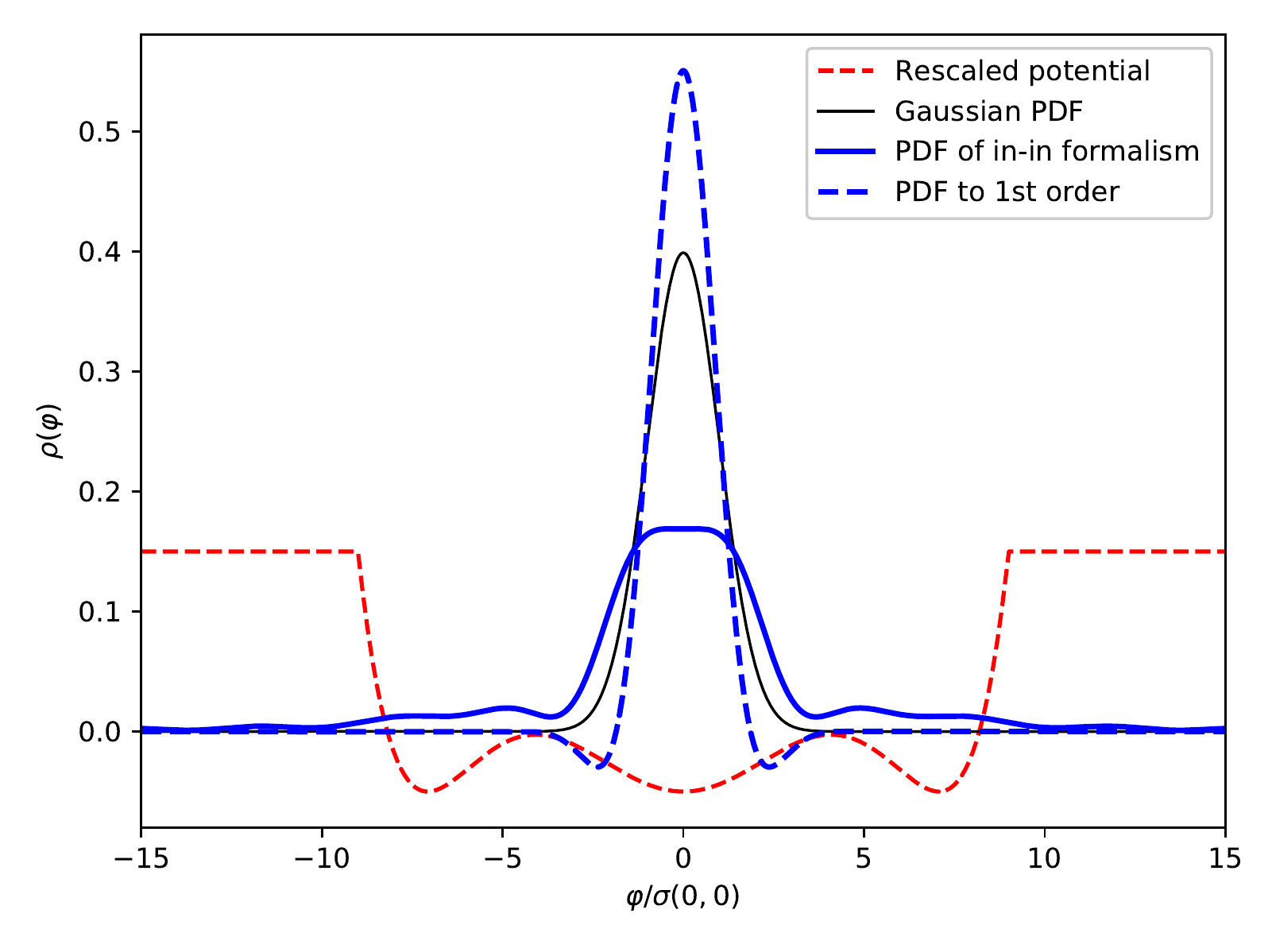}
\caption{The figure illustrates a typical PDF (solid blue) that may be obtained exactly using the \textit{in-in} formalism and the tools developed herein, in the non-perturbative regime $A = 1$ of~\eqref{V-example}. A Gaussian PDF (solid black) and the first order result in perturbation theory (dashed blue) are shown for comparison. The PDF was generated with $\sigma^2(x,0) = (1/5 - 4i/5) \sigma^2(0,0) $. The potential that gives rise to the non-Gaussian deformations is also displayed for comparison (dashed red). Its amplitude is not to scale, but the field range has been rescaled by ${\rm Re}\{\sigma^2(x,0)\}/\sigma^2(0,0)$ in order to properly account for the transfer.}
\label{fig:FIG_2}
\end{figure}

Note that the covariance matrix is nonsingular only because $\sigma^2(x,0) \neq \sigma^2(0,x)$, and consequently it distinguishes the fields $\varphi_{\pm}$. If those covariances were equal, the Gaussian measure would reduce to a Dirac delta between $\varphi_+$ and $\varphi_-$ (thus cancelling the effects of the interaction $\mathcal{V}$) times a 2-D Gaussian in field space between $\varphi$ and $\varphi_\pm$. Then the integral over the auxiliary field variables $\varphi_{\pm}$ could be carried out explicitly yielding a pure Gaussian PDF for $\rho_\varphi(\x)$. Therefore, this tells us that $\rho_\varphi(\x)$ is nontrivial only if $\sigma^2(x,0) = \sigma^2(0,x)^*$ has a nonzero imaginary part. As we will remark later in section~\ref{sec:conclusions}, this is the manner in which causality arises within the \textit{in-in} formalism: the boundary of the spacetime region where the imaginary part of $\sigma^2(x,0)$ is nonzero draws the light cone of the particles associated to the field. The influence of different backgrounds is thus taken into account through the construction of the propagator. 

Now the result is non-perturbative, and we need not concern ourselves with the convergence of sums/integrals. Thus, in principle, we may make a choice on the function $\mathcal{V}(\varphi)$ as we see fit. To illustrate our point in choosing a nontrivial structure, let
\be \label{V-example}
 \mathcal{V}(\varphi) = A \min\{5 ,\varphi^6 - 4\varphi^4 + 4\varphi^2 \}
\ee
in units where the free theory variance of the field satisfies $\sigma^2(0,0)=1$, and $A$ is a dimensionless real parameter. This is nothing but a local potential well in field space with features inside it. Furthermore, we take the theory to be such that $\sigma^2(0,0) =  \sigma^2(x,x)$ for all spacetime positions $x$. 

Having made these choices, we may now carry out the integrations in~\eqref{0dim-PDF} numerically for a fixed value of $\sigma^2(x,0)$ and see how the effects of such a potential propagate in spacetime. The result of this procedure is shown in Figures~\ref{fig:FIG_1} and~\ref{fig:FIG_2} for two different values of the parameter $A$: 0.1 (presumably perturbative) and 1 (non-perturbative), so as to compare with our previous result~\eqref{rpsi2} and assess its validity. Since the real part of the covariance $\sigma^2(x,0)$ is smaller than the typical field range $\sigma^2(0,0)$, we have to rescale the argument of the potential when presenting the plots in order to make meaningful comparisons involving the transfer of statistics from the singular event in spacetime to the field. This is because the external field variable $\varphi$ will appear in~\eqref{0dim-PDF} in specific combinations, such as $(\varphi_+ - \varphi \, \sigma^2(x,0)/\sigma^2(0,0))$.

Figure~\ref{fig:FIG_1} shows the results of taking the proposed potential with $A = 0.1$. Although the PDF to 1st order is distinguishable from the exact result, they qualitatively agree in the central region $|\varphi/\sigma(0,0)| \leq 3$: both have an increase in probability around $\varphi \sim 0$ and a slight decrease in the range $1 \leq |\varphi/\sigma(0,0)| \leq 3$ with respect to the original Gaussian distribution. The difference between them is that the exact result is able to account for tunneling to the minima of the rescaled potential $\mathcal{V}(\varphi {\rm Re}\{\sigma^2(x,0)\}/\sigma^2(0,0))$ at $\varphi/\sigma(0,0) \sim \pm 7$, while the perturbative result offers negligible probability in comparison. This, in turn, forces the 1st order result to accommodate more probability in the central region $\varphi \sim 0$, thus explaining its overestimation.

On the other hand, Figure~\ref{fig:FIG_2} displays the results of taking the same setting, but with $A = 1$. In this situation it is clear that the perturbative result breaks down, as the PDF to 1st order takes negative values around $|\varphi/\sigma(0,0)| \sim 3$ and wildly overestimates the value of the probability density around $\varphi \sim 0$. Conversely, the exact result obeys a more complex behavior than before: the PDF spreads out more profusely, giving a non-negligible probability to the minima at $\varphi/\sigma(0,0) \sim \pm 7$, but still retaining its central maximum at $\varphi \sim 0$. Moreover, there seem to be small oscillatory features that go beyond the range of the potential, which probably a perturbative expansion would have a hard time describing at any finite order.

Let us emphasize that, although simplistic, this type of self-interactions $V(\varphi,\r,t) = \delta(t) \delta^{(3)}(\r) \mathcal{V} (\varphi)$ may serve to describe any non-trivial feature in field space, provided it is localized in spacetime and the observer is sufficiently far away from the source of these signals, so that the covariance $\sigma^2(x,0)$ does not vary significantly when scanning the region of interest. Indeed, if one manages to incorporate homogeneity and isotropy into these results, they may serve as a toy model in the search of primordial non-Gaussianity in our universe.

As a final comment on this section, let us note that equation~\eqref{0dim-PDF} could have been derived directly from a path integral approach. However, we found it to be more transparent regarding the structure of the theory to perform the resummation explicitly, as this is what one would have to do in a general setting, after (hopefully) having solved some of the spatial or temporal integrations.

Now we turn to an example where the field integrals may be solved at once, but the spacetime integrals will still remain in the result.

\subsection{A reducible example with the in-out formalism}

Consider a real scalar field $\varphi$, with a free theory given by a known quadratic Lagrangian $\mathcal{L}_{\text{free}}$, in the presence of a self-interaction $V(\varphi) = \Lambda^4 e^{\varphi/f}$. In order to simplify the discussion involving the diagrammatical ``external legs'', we will take the \textit{in} and \textit{out} states to be excitations around the free theory vacuum, instead of the true vacuum of the theory (where $V(\varphi)$ plays a role). This could be implemented through a spacetime dependence of the parameter $f$, or by simply setting $\Lambda = 0$ towards the infinities of spacetime, so to justify the absence of the potential's effects in this regard. However, to alleviate the notation we shall omit these dependencies and regard them as constants. Starting from~\eqref{in-out-biPDF} and taking time-ordered propagators $\sigma_T^2(x_1 - x_2)$, the amplitude $\braket{f|i}$ (without removing disconnected diagrams) reads
\begin{widetext}
\be
\braket{f|i} = \sum_{N = 0}^{\infty} \frac{(-i \Lambda^4)^N}{N!} \int_{x_1} ... \int_{x_N} \int_{\varphi_{\rm ext}} \int_{\varphi_{\rm int}} f(\varphi_{\rm ext}) \, \rho_G(\varphi_{\rm ext}, \varphi_{\rm int}; \Sigma_T) \,  e^{(\varphi_1 + \cdots + \varphi_N)/f},
\ee
\end{widetext}
where $\int_{x} = \int_{-\infty}^{\infty} dt \int_{\x}$ is an integral over all of spacetime, $\int_{\varphi_{\rm ext}}$ and $\int_{\varphi_{\rm int}}$ denote integrals over the value of the field at given spacetime positions, which are set by hand in the case of $\varphi_{\rm ext}$ and are set by the integration coordinates $(x_1,...,x_N)$ for $\varphi_{\rm int} = (\varphi_1, ..., \varphi_N)$. $f(\varphi_{\rm ext})$ is a function of the ``external'' fields only, and represents the observable transition amplitude of interest. Finally, $\rho_G$ represents a multivariate Gaussian probability density involving both $\varphi_{\rm ext}$ and $\varphi_{\rm int}$ characterized by a covariance matrix $\Sigma_T$ composed by time-ordered propagators $\sigma_T(x_1,x_2)$ relating the fields' corresponding spacetime positions. The fields that define the \textit{in} state have their coordinates evaluated as if they were at $t = -\infty$, and correspondingly at $t = +\infty$ for the fields that define the \textit{out} state.
 
In the case of this specific potential, it is possible to carry out the ``internal'' field integrations (i.e. those that account for quantum corrections) to obtain
\begin{widetext}
\be
\braket{f|i} = \sum_{N = 0}^{\infty} \frac{(-i \Lambda^4)^N}{N!} \int_{x_1} ... \int_{x_N} \int_{\varphi_{\rm ext}} \rho_G(\varphi_{\rm ext}) \, f \! \left(\varphi_{\rm ext} + \frac{1}{f} \sum_{i=1}^N \sigma_T^2(x_i - x_{\rm ext}) \right) \exp \left( \frac{1}{2f^2} \sum_{i,j=1}^{N} \sigma_T^2(x_i - x_j) \right)
\ee
and
\be
\braket{0|S|0} = \sum_{N = 0}^{\infty} \frac{(-i \Lambda^4)^N}{N!} \int_{x_1} ... \int_{x_N} \exp \left( \frac{1}{2f^2} \sum_{i,j=1}^{N} \sigma_T^2(x_i - x_j) \right).
\ee
\end{widetext}
Now one may ask about the expectation value for a given function $f(\varphi_{\rm ext})$. The first object that comes to mind is a propagator (which is not exactly the case, since the vacuum $\ket{0}$ precludes the effects of the potential). Let us take $f(\varphi_{\rm ext}) = \varphi(r_1)|_{t=-\infty} \, \varphi(r_2)|_{t=+\infty}$, where the subscripts $\pm \infty$ inform us on how the time-ordering must be taken within the covariances. After a few manipulations, one can show the result to be
\begin{widetext}
\be \label{propagator-exp}
\begin{split}
\frac{\braket{f|i}}{\braket{0|S|0}} = \sigma^2(r_1 - r_2) + \frac{1}{f^2 \braket{0|S|0}} \sum_{N = 0}^{\infty} \frac{(-i \Lambda^4)^N}{N!} & \int_{x_1} ... \int_{x_N} \left(N \sigma_T^2(r_1 - x_1) \sigma_T^2(x_1 - r_2) \right. \\ + & \left. N (N-1) \sigma_T^2(r_1 - x_1) \sigma_T^2(x_2 - r_2) \right) \exp \left( \frac{1}{2f^2} \sum_{i,j=1}^{N} \sigma_T^2(x_i - x_j) \right).
\end{split}
\ee
\end{widetext}
This equation has a clean interpretation in terms of the usual diagrammatic expansions: the first term inside the parenthesis that is being integrated represents the two external states connected to a single vertex, thus leaving no external momenta to flow through the rest of the diagram, while the second term connects each state \textit{in} and \textit{out} to a different vertex, signalling a different type of diagram where external momentum will flow through it and yield nontrivial ``loops''. However, they have all been re-summed into a single exponential, which plays the role of an interaction kernel.

This structure may be of use to deal with other interactions involving an exponential of the fields. For instance, in a model with two scalar fields, a coupling alike
\be
V(\varphi,\phi) = \sum_{n=0}^{\infty} \frac{\varphi^n (\hat{d} \cdot \nabla)^n}{f^n n!} \phi = e^{\varphi \hat{d} \cdot \vec{\nabla}/f } \phi,
\ee
where $\vec{\nabla}$ instructs the gradient to act only towards the right (i.e. only on $\phi$), should follow essentially the same structure as equation~\eqref{propagator-exp} in the $\varphi$ ``loops'', while retaining a quasi-quadratic structure on $\phi$.

A final note can be made in that~\eqref{propagator-exp} can be regarded as non-perturbative in terms of $1/f$ at each order in $\Lambda^4$, because no series expansion was ever performed in terms of this parameter to obtain the result. This occurs in the same way as the structure of the potential is preserved in the first-order results~\eqref{rpsi2} and~\eqref{k-PDF}, where whatever ``intrinsic'' parameter that plays no (or little) role in determining the amplitude of the interaction is, in principle, kept to all orders.

\section{Discussion \& Conclusions} \label{sec:conclusions}

We have studied how probability distribution functionals arise perturbatively within a large class of quantum field theories, and derived general formulae that allow one to compute these distributions at every order in perturbation theory starting from a Hamiltonian approach. If done carefully, the full result may be summed back into a functional integral that can be readily connected to the path integral formulation of quantum mechanics; specifically to that of the in-in formalism~\cite{Weinberg:2005vy}. 

In what follows, we proceed to discuss some aspects of our results that may have been underemphasized earlier on.

\subsection{A note on causality}

As was pointed out in section~\ref{sec:example}, the PDF resulting from considering a real covariance $\sigma^2$ is trivial (Gaussian). This is actually a general feature of the formalism: consider equation~\eqref{PDF}, take all the covariances to be real, and set a finite $t_0$ for definiteness, so that the $\epsilon$ prescription plays no role in selecting the \textit{in} state. Thus, as in this case it is possible to make permutations in the temporal arguments of $\sigma^2(t,t',r)$ without consequence in the result, what remains is, to every order in perturbation theory, a sum over the possible time-ordered volumes of integration with alternating signs. This sum vanishes identically.

Consequently, if we fix the $\r_i$ coordinates and integrate over time first, the result can only be non-vanishing if at least one covariance $\sigma^2(t,t',r)$ has a complex component. Thus, in practice the integrations in~\eqref{PDF} could be taken over a restricted region of spacetime. This region of spacetime depends on the number of vertices $V(\varphi)$ and the positions of the external fields $\x_i$, $\y_i$ in the amplitude. 

The preceding statement is closely related to the notion of causality. For instance, a massless scalar field can be shown to have a complex covariance matrix only if the arguments lie precisely on the light cone. Similarly, a massive scalar field has a complex covariance within the causal interior of the light cone. This should be true in general, with its exact behavior depending on the background metric of the problem at hand.

\subsection{On the nature of the possible infinities}

Typically, a perturbative expansion of a nonlinear theory will give rise to an asymptotic series with null convergence radius. However, as it should become obvious from Figure 1, this does not mean that the theory itself is divergent. Moreover, the equations that led to that result suggest that the proper way to compute observables is by leaving the field integrals (i.e. the path integral) outside the sum of the perturbative expansion. In a sense, it is reminiscent of the original formulation of the problem: at the beginning of the computation, we had to deal with objects of the form $\braket{\Psi | \mathcal{O} | \Psi}$, where the inner product (``braket'') implies effecting the contractions between the various field operators that appear. This is exactly the role played by the field integrals with respect to a Gaussian measure, thus suggesting that the precise expressions for the PDF are with these integrals left outside the perturbative expansion, where ostensibly no perturbative divergence should be encountered.

Thus, the infinities that may still roam within the theory are those induced from the linear theory: namely, the exact expression for the propagator $\sigma^2(x_1,x_2)$ usually implies divergences in the correlation functions when computing loop diagrams. These are the ones usually tackled by regularization schemes. As they may be traced back to the free theory, they have no consequence in the results presented herein and may be addressed through the procedure most akin to the computation that is being undertaken.

\subsection{Concluding remarks}

Perhaps the most appealing feature of the results contained herein is that they circumvent the need to compute diagrams individually by considering the complete interaction $V(\varphi)$ instead of expanding it in a Taylor series and treating each term separately. This can prove crucial to the tractability of theories with a rich structure in the self-interactions, as is the case in~\cite{Chen:2018uul, Chen:2018brw}. In this type of situations, the formulae presented herein may be particularly useful because it is usually possible to solve at least one of the integrals analytically and then sum back the result into a more compact object.

Derivative couplings can also be treated in this manner, although in this case additional propagators need to be defined, and additional Gaussian integrals are required to implement them. In this scheme, the field $\varphi$ and its derivatives $\partial_{\mu} \varphi$ within an interacting term $V$ would be treated as independent variables, as each of them would be running over all their possible values when computing the relevant amplitudes. The pith of the matter is that because we only require to know the free theory propagators to describe the interacting theory perturbatively, in this regime it should always be possible to describe the full probability distributions in terms of conditional expectation values deriving from a Gaussian PDF.

On a different note, throughout this article we assumed mode functions $\varphi_k(t)$ dependent only on the absolute value of the momentum, thus giving a position-space two-point correlation that, disregarding the temporal arguments, is only a function of the distance between the two points. However, this is not a necessary condition: provided a solution for the free theory defined through a two-point function $\sigma^2(x_1, x_2)$ and Wick's theorem, most of the results presented herein hold true. Theories with a naturally arising inhomogeneous background, violating momentum conservation, or those with an anisotropic medium, having different signal propagation speeds in each direction, also have probability densities of the same form as those presented in this work.

All in all, we expect these results to find applications in a vast variety of settings, including (but not exclusive to) condensed matter and cosmology, where the use of quantum field theory is an ubiquitous necessity for the computation of observable quantities. 

\begin{acknowledgments}

We thank Jorge Alfaro, Rafael Bravo, Luis E. F. Fo\`{a} Torres, Fernando Lund, Gonzalo Palma, and Spyros Sypsas for useful discussions. BSH is supported by a CONICYT grant number CONICYT-PFCHA/Mag\'{i}sterNacional/2018-22181513, and also acknowledges support from the Fondecyt Regular project number 1171811 (CONICYT).

\end{acknowledgments}

\begin{appendix}

\section{The loop structure within the $n$-point functions} \label{sec:loops}

For the sake of familiarity with traditional approaches to QFT, we will proceed with the computation mostly in momentum space, even though the final result will reveal this step as unnecessary.

Let $\int_{\gamma} = (2\pi)^{-1} \int_{-\infty}^\infty d\gamma$ and $\int_{\varphi} = \int_{-\infty}^\infty d\varphi$. Then replacing equations~\eqref{inter} and~\eqref{expansion} into~\eqref{npoint}, we get
\begin{widetext}
\be \label{npoint2}
\begin{split}
\int_{\r_1} \int_{\varphi_1} \int_{\gamma_1} ... \int_{\r_N} \int_{\varphi_N} \int_{\gamma_N} V(\varphi_1, \r_1, t_1) e^{i \gamma_1 \varphi_1} ... V(\varphi_N, \r_N, t_N) e^{i \gamma_N \varphi_N} \sum_{m_1 = 0}^{\infty} ... \sum_{m_N = 0}^{\infty} \frac{(-i \gamma_1)^{m_1} ... (-i \gamma_N)^{m_N}}{m_1! ... m_N!} \int_{k_{11}} ... \int_{k_{1m_1}} \\ 
... \int_{k_{N1}} ... \int_{k_{Nm_N}}  \bra{0} \! \hat{\varphi}_I(\q_1, t_0) ... \hat{\varphi}_I(\q_J, t_0) \hat{\varphi}_I(\k_{11}, t_1) ... \hat{\varphi}_I(\k_{1m_1}, t_1) ... \hat{\varphi}_I(\k_{l1}, t_l) ... \hat{\varphi}_I(\k_{lm_l}, t_l)  \hat{\varphi}_I(\k_1,t) ... \hat{\varphi}_I(\k_n,t) \\
\hat{\varphi}_I(\k_{(l+1)1}, t_{l+1}) ... \hat{\varphi}_I(\k_{(l+1)m_{l+1}}, t_{l+1}) ... \hat{\varphi}_I(\k_{N1}, t_N) ... \hat{\varphi}_I(\k_{Nm_N}, t_N) \hat{\varphi}_I(\p_1, t_0) ... \hat{\varphi}_I(\p_J, t_0) \! \ket{0} \prod_{j=1}^N e^{-i \sum_{a=1}^{m_j} \k_{ja} \cdot \r_j}.
\end{split}
\ee
\end{widetext}
Let us not get distracted by the size of the previous equation and instead focus on how to deal with it. The previous vacuum expectation value can be evaluated by moving all annihilation operators to the right, giving rise to contractions between pairs of field operators in all possible ways. Hence, we need only distinguish the non-equivalent pairings and count the number of equivalent contractions for each pairing. 

Now we take a diagrammatic approach and try to obtain the fully interacting contributions. That is, in what follows we will only keep track of the terms where all fields at times \(t_0\) or \(t\) are contracted with fields arising from the interaction-picture hamiltonian. We regard two contractions as equivalent if they are connected to the same pair of spacetime positions (or vertices), indexed by the letter \(l\). Additionally, we define $n_{ij}$ as the number of field contractions between vertices $i$ and $j$, thus making $n_{ii}$ the number of closed loops formed from vertex $i$ alone. Also, we define $n_i$ as the number of contractions from vertex $i$ to the outer fields in the correlation (i.e., the ones evaluated at time $t$ or $t_0$). 

Firstly, let us count the number of possible ways of assigning roles to each field in the correlation: at vertex $i$, we have that the following multinomial coefficient
\be
\binom{m_i}{n_i, 2n_{ii}, n_{i1}, ..., n_{iN}}
\ee
is the number of possible ways of assigning the fields of vertex $i$ to the different roles they can undertake, with 
\be
m_i =  n_i + 2n_{ii} + \sum_{j \neq i} n_{ij} 
\ee
referring to the indices in the power series of~\eqref{npoint2} that represents $e^{-i\gamma_i \hat \varphi}$. Once these roles have been assigned, we may count the number of equivalent ways to achieve a certain configuration of contractions: given $n_{ij}$, if $i \neq j$, there are $n_{ij}!$ ways of forming contractions between vertices $i$ and $j$, and similarly there are
\be
\frac{(2n_{ii})!}{2^{n_{ii}} n_{ii}!}
\ee
ways of arranging the contractions of vertex $i$ with itself. Since the ``outer legs'' of~\eqref{npoint2} (the fields with momenta $\p_i$ or $\k_i$) are distinguishable, the only remaining combinatorial factor to account for is $n_i!$, which is the number of possible ways of assigning the $n_i$ ``outer legs'' to the $n_i$ fields of the vertex available for these contractions.

With all the previous statements considered,~\eqref{npoint2} is equal to
\begin{widetext}
\be \label{npoint3}
\begin{split}
\int_{\r_1} \! \int_{\varphi_1} \! \int_{\gamma_1} \! ... \! \int_{\r_N} \! \int_{\varphi_N} \! \int_{\gamma_N} \! V(\varphi_1, \r_1, t_1) e^{i \gamma_1 \varphi_1} ... V(\varphi_N, \r_N, t_N) e^{i \gamma_N \varphi_N } \!\!\underbrace{\sum_{n_1 = 0}^{n+J} ... \sum_{n_N = 0}^{n+J}}_{n_1 + ... + n_N = n + J}  \sum_{n_{11} = 0}^{\infty} \sum_{n_{12} = 0}^{\infty} ... \! \sum_{n_{1N} = 0}^{\infty}  \sum_{n_{22} = 0}^{\infty} \sum_{n_{23} = 0}^{\infty} ... \! \sum_{n_{NN} = 0}^{\infty} \\
\left[ \frac{(-i \gamma_1)^{ n_1 + 2n_{11} + \sum_{j \neq 1} n_{1j} } ... (-i \gamma_N)^{n_N  + 2n_{NN} + \sum_{j \neq N} n_{Nj}  }}{\left( n_1 + 2n_{11} + \sum_{j \neq 1} n_{1j}  \right)! \, ... \, \left(n_N + 2n_{NN} + \sum_{j \neq N} n_{Nj}  \right)!} \binom{ n_1 + 2n_{11} + \sum_{j \neq 1} n_{1j}  }{n_1, 2n_{11}, n_{11}, ..., n_{1N}} ... \binom{ n_N + 2n_{NN} + \sum_{j \neq N} n_{Nj}  }{n_N, 2n_{NN}, n_{N1}, ..., n_{NN}} \right. \\
\left. n_1! ... n_N!  \frac{(2n_{11})!}{2^{n_{11}} n_{11}!} \left( \int_k \Delta(t_1, t_1, k) \right)^{n_{11}} \!\!\!\! n_{12}! \left( \int_k \Delta(t_1, t_2, k) e^{i \k \cdot (\r_1 - \r_2)}\right)^{n_{12}}  ... \,  \frac{(2n_{NN})!}{2^{n_{NN}} n_{NN}!} \left( \int_k \Delta(t_N, t_N, k) \right)^{n_{NN}}  \right] \\
\sum_{\{i_b\}} \exp\left(-i \sum_{b=1}^{n+2J} \q_{b} \cdot \r_{i_b}\right) \left( \Delta(t_0, t_{i_1}, q_1) ... \Delta(t_0, t_{i_{J}}, q_J) \Delta(t_{i_{J+1}}, t, k_1)  ... \Delta(t, t_{i_{J+n}}, k_n) \Delta(t_{i_{J+n+1}}, t_0, p_1) ... \Delta(t_{i_{n+2J}}, t_0, p_J) \right)
\end{split}
\ee
\end{widetext}
where we have written $\q_{b} = \k_{b-J}$ for $b = J+1, ..., J+n$ and $\q_b = \p_{b-(J+n)}$ for $b=J+n+1, ..., 2J+N$ in the exponential of the last line. This last sum (over $\{i_b\}$) accounts for all possible ways of connecting the outer legs to the vertices, with the restriction that $i_b$ must take the value $a$ for $n_a$ values of $b$. On the other hand, a myriad of cancellations occur inside the square bracket. After carrying them out, we obtain:
\begin{widetext}
\be \label{npoint4}
\begin{split}
\int_{\r_1} \! \int_{\varphi_1} \! \int_{\gamma_1} \! ... \! \int_{\r_N} \! \int_{\varphi_N} \! \int_{\gamma_N} \! V(\varphi_1, \r_1, t_1) e^{i \gamma_1 \varphi_1} ... V(\varphi_N, \r_N, t_N) e^{i \gamma_N \varphi_N } \!\!\underbrace{\sum_{n_1 = 0}^{n+J} ... \sum_{n_N = 0}^{n+J}}_{n_1 + ... + n_N = n + J}  \sum_{n_{11} = 0}^{\infty} \sum_{n_{12} = 0}^{\infty} ... \! \sum_{n_{1N} = 0}^{\infty}  \sum_{n_{22} = 0}^{\infty} \sum_{n_{23} = 0}^{\infty} ... \! \sum_{n_{NN} = 0}^{\infty} \\
(-i \gamma_1)^{n_1} ... (-i \gamma_N)^{n_N} \left[  \left( \prod_{i<j}^N \frac{1}{n_{ij}!} \left( - \gamma_i \gamma_j \int_k \Delta(t_i, t_j, k) e^{i \k \cdot (\r_i - \r_j)}\right)^{n_{ij}} \right) \left(\prod_{i=1}^N \frac{1}{n_{ii}!} \left( - \frac{\gamma_i^2}{2} \int_k \Delta(t_i, t_i, k) \right)^{n_{ii}}  \right) \right] \\
\sum_{\{i_b\}} \exp\left(-i \sum_{b=1}^{n+2J} \q_{b} \cdot \r_{i_b}\right) \left( \Delta(t_0, t_{i_1}, q_1) ... \Delta(t_0, t_{i_{J}}, q_J) \Delta(t_{i_{J+1}}, t, k_1)  ... \Delta(t, t_{i_{J+n}}, k_n) \Delta(t_{i_{J+n+1}}, t_0, p_1) ... \Delta(t_{i_{n+2J}}, t_0, p_J) \right)
\end{split}
\ee
which can be recast as
\be \label{npoint5}
\begin{split}
\underbrace{\sum_{n_1 = 0}^{n+2J} ... \sum_{n_N = 0}^{n+2J}}_{n_1 + ... + n_N = n + 2J} \!\! \int_{\r_1} \!\! \int_{\varphi_1} \!\! \int_{\gamma_1} \!\! ... \!\! \int_{\r_N} \!\! \int_{\varphi_N} \!\! \int_{\gamma_N} \! \dfrac{\partial^{n_1} V}{\partial \varphi_1^{n_1}} ... \dfrac{\partial^{n_N} V}{\partial \varphi_N^{n_N}} e^{i \sum_{j} \gamma_j \varphi_j} \exp \left( \sum_{i=1}^N \left( -\frac{1}{2} \gamma_i^2 \sigma_0^2(t_i) \right) + \sum_{\substack{i<j \\ i,j=1}}^N \left( - \gamma_i \gamma_j \sigma^2(t_i, t_j, |\r_i - \r_j|) \right) \right) \\
\sum_{\{i_b\}} \exp\left(-i \sum_{b=1}^{n+2J} \q_{b} \cdot \r_{i_b}\right) \left( \Delta(t_0, t_{i_1}, q_1) ... \Delta(t_0, t_{i_{J}}, q_J) \Delta(t_{i_{J+1}}, t, k_1)  ... \Delta(t, t_{i_{J+n}}, k_n) \Delta(t_{i_{J+n+1}}, t_0, p_1) ... \Delta(t_{i_{n+2J}}, t_0, p_J) \right)
\end{split}
\ee
where we have omitted the arguments of the potential $V(\varphi, \r, t)$. Now performing the integrations over the \(\gamma\) variables, we end up with
\be \label{npoint55}
\begin{split}
\underbrace{\sum_{n_1 = 0}^{n+2J} ... \sum_{n_N = 0}^{n+2J}}_{n_1 + ... + n_N = n + 2J} \int_{\r_1} \int_{\varphi_1} ... \int_{\r_N} \int_{\varphi_N} \dfrac{\partial^{n_1} V}{\partial \varphi_1^{n_1}} ... \dfrac{\partial^{n_N} V}{\partial \varphi_N^{n_N}}  \frac{\exp \left( -\frac{1}{2} \varphi_i \left(\Sigma_I^{-1}\right)_{ij} \varphi_j \right)}{\sqrt{(2\pi)^N |\text{det} \Sigma_I|}}
\sum_{\{i_b\}} \exp\left(-i \sum_{b=1}^{n+2J} \q_{b} \cdot \r_{i_b}\right) \\ \left( \Delta(t_0, t_{i_1}, q_1) ... \Delta(t_0, t_{i_{J}}, q_J) \Delta(t_{i_{J+1}}, t, k_1)  ... \Delta(t, t_{i_{J+n}}, k_n) \Delta(t_{i_{J+n+1}}, t_0, p_1) ... \Delta(t_{i_{n+2J}}, t_0, p_J) \right)
\end{split}
\ee
\end{widetext}
where the matrix elements of $\Sigma_I$ are given by $(\Sigma_I)_{ij} = \sigma^2(t_{\text{min}\{i,j\}},t_{\text{max}\{i,j\}}, |\r_i -\r_j|)$, and we sum over repeated indices at this instance. Taking Fourier transform to position space over $\q_i$ (with conjugate variables $\z_i$), $\k_i$ (with conjugate variables $\x_i$) and $\p_i$ (with conjugate variables $\y_i$), we arrive at
\begin{widetext}
\be \label{npoint6a}
\begin{split}
\bra{0} \! \varphi_I(\z_1, t_0) ... \varphi_I(\z_J, t_0) H_I(t_1) ... H_I(t_l)   \varphi_I(\x_1,t) ... \varphi_I(\x_n,t) H_I(t_{l+1}) ... H_I(t_N) \varphi_I(\y_1, t_0) ... \varphi_I(\y_J, t_0) \! \ket{0}_{FI} \\
= \underbrace{\sum_{n_1 = 0}^{n+2J} ... \sum_{n_N = 0}^{n+2J}}_{n_1 + ... + n_N = n + 2J} \int_{\r_1} \int_{\varphi_1} ... \int_{\r_N} \int_{\varphi_N} \dfrac{\partial^{n_1} V}{\partial \varphi_1^{n_1}} ... \dfrac{\partial^{n_N} V}{\partial \varphi_N^{n_N}}  \frac{\exp \left( -\frac{1}{2} \varphi_i \left(\Sigma_I^{-1}\right)_{ij} \varphi_j \right)}{\sqrt{(2\pi)^N |\text{det} \Sigma_I|}} \\
\sum_{\{i_b\}} \left( \sigma^2(t_0, t_{i_1}, |\z_1 - \r_{i_1}|) ... \sigma^2(t_0, t_{i_{J}}, |\z_J - \r_{i_J}|) \sigma^2(t_{i_{J+1}}, t, |\x_1 - \r_{i_{J+1}}|) \right. \\  \left. ... \sigma^2(t, t_{i_{J+n}}, |\x_n - \r_{i_{J+n}}|) \sigma^2(t_{i_{J+n+1}}, t_0, |\y_1 - \r_{i_{J+n+1}}|) ... \sigma^2(t_{i_{n+2J}}, t_0, |\y_J - \r_{i_{n+2J}}|) \right).
\end{split}
\ee
\end{widetext}
So far, we have only dealt with the fully interacting contributions to the correlation. But from this end of the computation, we can appreciate some structure emerging in the result: it is an expectation value over a gaussian probability density function. With this in mind, we claim that we can write down
\begin{widetext}
\be \label{npoint7}
\begin{split}
\bra{0} \! \varphi_I(\z_1, t_0) ... \varphi_I(\z_J, t_0) H_I(t_1) ... H_I(t_l)   \varphi_I(\x_1,t) ... \varphi_I(\x_n,t) H_I(t_{l+1}) ... H_I(t_N) \varphi_I(\y_1, t_0) ... \varphi_I(\y_J, t_0) \! \ket{0} = \int_{\varphi_{\z_1}} \!\!\!\!\!\! ... \! \int_{\varphi_{\z_J}}  \\
\int_{\varphi_{\x_1}} \!\!\!\!\!\!  ... \! \int_{\varphi_{\x_N}} \!\int_{\varphi_{\y_1}}  \!\!\!\!\!\! ... \! \int_{\varphi_{\y_J}} \! \int_{\r_1} \int_{\varphi_1} ... \int_{\r_N} \int_{\varphi_N} V(\varphi_1, \r_1, t_1) ...  V(\varphi_N, \r_N, t_N)  \frac{\exp \left( -\frac{1}{2} \pmb{\varphi}^T \cdot \left(\pmb{\Sigma}^{-1}\right) \cdot \pmb{\varphi} \right)}{\sqrt{(2\pi)^{N+n+2J} |\text{det} \pmb{\Sigma} |}} \varphi_{\z_1} ... \varphi_{\z_J} \varphi_{\x_1} ... \varphi_{\x_n} \varphi_{\y_1} ... \varphi_{\y_J} 
\end{split}
\ee
with \(\pmb{\varphi}^T \equiv (\varphi_{\z_1} \,\,\, ... \,\,\, \varphi_{\z_J}  \,\,\, \varphi_{1} \,\,\, ... \,\,\, \varphi_l \,\,\, \varphi_{\x_1} \,\,\, ... \,\,\, \varphi_{\x_n} \,\,\, \varphi_{l+1} \,\,\, ... \,\,\, \varphi_N \,\,\, \varphi_{\y_1} \,\,\, ... \,\,\, \varphi_{\y_J}  )\), and \(\pmb{\Sigma}\) as the corresponding covariance matrix. 
\end{widetext} 

The covariances in the last expression are the propagators between the fields' corresponding spacetime positions, with their respective temporal arguments ordered as the fields are in the definition of $\pmb{\varphi}^T$. For example, the covariance relating $\varphi_{\z_a}$ and $\varphi_{b}$ is $\sigma^2(t_0,t_b,|\z_a-\r_b|)$, and the one relating $\varphi_{\x_i}$ with $\varphi_{b}$ would be $\sigma^2(t,t_b,|\x_i-\r_b|)$ if $b\geq l+1$, while it would be $\sigma^2(t_b,t,|\x_i-\r_b|)$ if $b \leq l$.  

Note that the part of the correlations which we haven't computed explicitly in this appendix is given by terms that are products of free theory pairings between the external legs and a fully interacting contribution involving the remaining fields (it is not essential that the number of fields at $t_0$ is equal at both sides of the interaction), and it is fairly easy to check that the correlators of the free theory are given by a gaussian distribution as in~\eqref{npoint7} without the $\varphi_i$ terms. So, \textit{a posteriori}, the claim doesn't seem unreasonable. The proof is given in the main text.

\vspace{0.5cm}

\section{The PDF for an arbitrary initial state} \label{sec:arbitrary-initial}

Had we kept the fields that define the \textit{in} state within section~\ref{subsec:main-PDF}, we would have arrived to
\begin{widetext}
\be
\begin{split}
\langle \varphi(\x_1,t) ... \varphi(\x_n,t) \rangle =  \sum_{N=0}^{\infty} (-i)^N \sum_{l=0}^{N} (-1)^l \int_{t_0 - i\epsilon |t_0|}^t \!\!\!\!\!\!\!\!\! dt_l ...
\int_{t_0 - i\epsilon |t_0|}^{t_2} \!\!\!\!\!\!\!\!\! dt_1 \int_{t_0 + i\epsilon |t_0|}^t \!\!\!\!\!\!\!\!\! dt_{l+1} ... \int_{t_0 + i\epsilon |t_0|}^{t_{N-1}} \!\!\!\!\!\!\!\!\! dt_N \int_{\r_1} ... \int_{\r_N} \!\int_{\varphi_1} \!\!\!  ... \int_{\varphi_N}  \\ \int_{\varphi_{\z_1}} \!\!\!\!\!\! ... \! \int_{\varphi_{\z_J}} \!
\int_{\varphi_{\x_1}} \!\!\!\!\!\!  ... \! \int_{\varphi_{\x_N}} \! \int_{\varphi_{\y_1}}  \!\!\!\!\!\! ... \! \int_{\varphi_{\y_J}} V(\varphi_1, \r_1, t_1) ...  V(\varphi_N, \r_N, t_N)  \frac{\exp \left( -\frac{1}{2} \pmb{\varphi}^T \cdot \left(\pmb{\Sigma}^{-1}\right) \cdot \pmb{\varphi} \right)}{\sqrt{(2\pi)^{N+n} |\text{det} \pmb{\Sigma} |}} \varphi_{\z_1} ... \varphi_{\z_J} \varphi_{\x_1} ... \varphi_{\x_n} \varphi_{\y_1} ... \varphi_{\y_J} 
\end{split}
\ee
\end{widetext}
where we haven't set $\z_i = \y_i$ yet in order to avoid equivocal statements, since the covariance associated to a contraction of $\varphi_\z$ with a vertex is not equal to that of $\varphi_\y$ (in fact they are complex conjugates). 

The subsequent steps follow in the same way as in the main text, yielding
\begin{widetext}
\be \label{PDFalt}
\begin{split}
\rho_{\varphi} = \sum_{N=0}^{\infty} (-i)^N \sum_{l=0}^{N} (-1)^l \int_{t_0 - i\epsilon |t_0|}^t \!\!\!\!\!\!\!\!\! dt_l ...
\int_{t_0 - i\epsilon |t_0|}^{t_2} \!\!\!\!\!\!\!\!\! dt_1 \int_{t_0 + i\epsilon |t_0|}^t \!\!\!\!\!\!\!\!\! dt_{l+1} ... \int_{t_0 + i\epsilon |t_0|}^{t_{N-1}} \!\!\!\!\!\!\!\!\! dt_N \int_{\r_1} ... \int_{\r_N} \!\int_{\varphi_1} \!\!\!  ... \int_{\varphi_N} \\  \int_{\varphi_{\z_1}} \!\!\!\!\!\! ... \! \int_{\varphi_{\z_J}}
 \! \int_{\varphi_{\y_1}}  \!\!\!\!\!\! ... \! \int_{\varphi_{\y_J}}   \frac{\exp \left( -\frac{1}{2} \pmb{\varphi}^T \cdot \left(\pmb{\Sigma}^{-1}\right) \cdot \pmb{\varphi} \right)}{\sqrt{(2\pi)^{N+n} |\text{det} \pmb{\Sigma} |}} \varphi_{\z_1} ... \varphi_{\z_J} V(\varphi_1, \r_1, t_1) ...  V(\varphi_N, \r_N, t_N) \varphi_{\y_1} ... \varphi_{\y_J} \big|_{\y_i = \z_i} ,
\end{split}
\ee
\end{widetext}
where we have to stress that when performing the field integrals in this last expression the arguments of the covariances must be taken in the same order as the fields are written, and after doing that, set $\y_i = \z_i$ so that the \textit{in} states match. 

We can also write this as a functional integral
\begin{widetext}
\be \label{Path-integral2a}
\begin{split}
\rho_{\varphi} =& \int \!  D\varphi_- D\varphi_+ \!   \left( \varphi_-(\y_1,t_0) ... \varphi_-(\y_J,t_0) \exp \left\{ +i\int^{t}_{t_0 - i\epsilon |t_0|} \!\!\!\!\!\!\!\!\! d t' \int_{\r} V(\varphi_-(\r,t'),\r,t') \right\} \right) \! \\ & \times \frac{\exp \left( - \frac{1}{2} \varphi \cdot \left( {\Sigma}^{-1} \right) \cdot \varphi \right)}{\sqrt{|\text{det} (2\pi \Sigma) |}} \! \left( \exp \left\{ - i \int^{t}_{t_0 + i\epsilon |t_0|} \!\!\!\!\!\!\!\!\! d t' \int_{\r} V(\varphi_+(\r,t'),\r,t') \right\} \varphi_+(\y_1,t_0) ... \varphi_+(\y_J,t_0) \right),
\end{split}
\ee
\end{widetext}
in which the distinction between $\varphi_+$ and $\varphi_-$ fields, with their respective time orderings, makes the result easier to write down. As in the main text, the resulting propagators/covariances involving a $\varphi_+$ field and another type of field ($\varphi_+$ or $\varphi$) always have the corresponding time in the second temporal argument, and are time-ordered if it is a $\varphi_+ \varphi_+$ contraction. Conversely, the entries of the propagators corresponding to $\varphi_-$ always go to the left when contracted with another type of field and are anti-time-ordered when considering a $\varphi_- \varphi_-$ contraction.

\end{appendix}

\twocolumngrid

\end{document}